\documentclass[11pt]{article}

\usepackage{amsmath,amssymb,mathtools}

\usepackage{amsthm}

\newtheorem{thrm}{Theorem}

\usepackage{times}
\usepackage{w-thm}
\usepackage[authoryear]{natbib}
\usepackage{xcolor}
\usepackage{graphicx,psfrag,epsf}
\usepackage{enumerate}
\usepackage{listings, mathtools}

\usepackage{tikz}
\usetikzlibrary{arrows,shapes,trees}
\usepackage{multirow,floatrow,enumitem,subfloat,float}
\usepackage{lineno,blindtext,setspace}
\usepackage{appendix,xr}
\usepackage{url}

\usepackage{setspace}
\usepackage{amscd}
\usepackage{latexsym}

\date{}

\def\b0{{0}}

\newcommand{\ts}{\texttt}

\graphicspath{{images/}} 

\usepackage[colorlinks,bookmarksopen,bookmarksnumbered,citecolor=red,urlcolor=red]{hyperref}

\newcommand\BibTeX{{\rmfamily B\kern-.05em \textsc{i\kern-.025em b}\kern-.08em
T\kern-.1667em\lower.7ex\hbox{E}\kern-.125emX}} 
\usepackage[margin=0.8in]{geometry}

\begin{document}

\begin{singlespace}
\title{SMARTp: A \underline{SMART} design for non-surgical treatments of chronic \underline{p}eriodontitis with spatially-referenced and non-randomly missing skewed outcomes}

\author{Jing Xu$^{1}$, Dipankar Bandyopadhyay$^{2}$, Sedigheh Mirzaei$^{3}$, Bryan Michalowicz$^{4}$, Bibhas Chakraaborty$^{5,}$\thanks{ \small{\textit{Address of Correspondence}: Centre for Quantitative Medicine and Programme in Health Services and Systems Research, Duke-NUS Medical School, Academia, 20 College Road, 06-41, Singapore, 169856. E-mail: \texttt{bibhas.chakraborty@duke-nus.edu.sg}}} \\ \\
\small{$^1$Cancer Data Science, Children's Medical Research Institute, The University of Sydney, Westmead, NSW, Australia} \\
\small{$^2$Department of Biostatistics, Virginia Commonwealth University}\\ 
\small{$^3$Department of Biostatistics, St. Jude Children's Research Hospital, Memphis, TN, USA}\\ 
\small{$^4$HealthPartners Institute, Bloomington, MN, USA}\\ 
\small{$^5$Centre for Quantitative Medicine and Programme in Health Services and Systems Research, Duke-NUS Medical School, Singapore}\\ 
}
\date{\ }
\maketitle
\end{singlespace}

\bigskip
\begin{singlespace}
\begin{abstract}
This paper proposes dynamic treatment regimes (DRTs) for choosing individualized effective treatment strategies of chronic periodontitis. The proposed DTRs are studied via \texttt{SMARTp} -- a two-stage sequential multiple assignment randomized trial (SMART) design. For this design, we propose a statistical analysis plan and a novel cluster-level sample size calculation method that factors in typical features of periodontal responses, such as non-Gaussianity, spatial clustering, and non-random missingness. Here, each patient/subject is viewed as a cluster, and a tooth within a subject's mouth is viewed as an individual unit inside the cluster, with the tooth-level covariance structure described by a conditionally autoregressive process. To accommodate possible skewness and tail behavior, the tooth-level clinical attachment level (CAL) response is assumed to be skew-$t$, with the non-randomly missing structure captured via a shared parameter model corresponding to the missingness indicator. The proposed method considers mean comparison for the regimes with or without sharing an initial treatment, where the expected values and corresponding variances or covariance for the sample means of a pair of DTRs are derived by the inverse probability weighting and method of moments. Simulation studies are conducted to investigate the finite-sample performance of the proposed sample size formula under a variety of outcome-generating scenarios. A major contribution of this work is the implementation of the sample size formula via a \texttt{R} package available in \texttt{GitHub}.

\vspace{12pt}
\end{abstract}
\end{singlespace}

\newpage

\section{Introduction}

Chronic periodontitis (CP) is a serious form of periodontal disease (PD), and if left untreated, continues to remain a major cause of adult tooth loss \citep{Eke_etal_2012}. It is a highly prevalent condition, affecting almost half of the US adults, $\ge 30$ years \citep{Thornton_etal_2013}. Periodontal disease (PD) maybe exhibit significant comorbidity, with diabetes, cardiovascular complications, respiratory illnesses, etc \citep{grossi1997treatment, wang2009periodontal, beck2001relationship}. Dental hygienists consider the \textbf{C}linical \textbf{A}ttachment \textbf{L}evel (or, CAL), rounded to the nearest millimeter, as the most important biomarker to measure the severity of PD \citep{Nicholls_2003}. The CAL refers to the amount of lost periodontal ligament fibers, with the severity categorized as Slight/Mild: 1-2mm, Moderate: 3-4mm, Severe: $\geq5$mm, according to the American Academy of Periodontology (AAP) 1999 guidelines \citep{wiebe2000periodontal}.

The treatment options for periodontitis range from oral hygiene, to scaling and root planing (SRP), to SRP with adjunctive treatments, and eventually to surgeries, when the severity increases. Recommended bi-annual (basic) dental cleaning, polishing, and professional flossing which we enjoy sitting in a comfortable reclining chair in a dental clinic only removes plaque (bacterial colony) and tartar above the gum line, and are not considered as effective procedures for treating gum diseases. Often, early stages of CP are effectively treated via non-surgical means. Benefits of non-surgical treatment of CP includes shorter recovery time due to less-invasive techniques, reduced discomfort than surgery, and fewer dietary restrictions leading to improved quality of life, post procedure.  When disease has progressed significantly, \textit{deep} cleaning, a two-step process consisting of scaling and (tooth) root planning (SRP), is often recommended as the gold standard \citep{herrera2016scaling} for thorough plaque removal, at or below the gum line. Yet, bacteria may still exist under the gum line following a SRP. Hence, the dentist or oral hygienist may also recommend various supplementary procedures, or adjuncts (such as locally-delivered antimicrobials, systemic antimicrobials, non-surgical lasers, etc) following SRP to treat chronically deep gum pockets. Also, patients with periodontal co-morbidity who responds sub-optimally to SRP may benefit from adjunctive therapy \citep{porteous2014adjunctive}. However, current evidence suggests that the use of these adjuncts as stand-alone treatments does not lead to any clinical benefits of treating CP, compared to SRP alone \citep{azarpazhooh2010effect, sgolastra2012efficacy}. Hence, they are usually recommended in conjunction to SRP.

In 2011, the Council on Scientific Affairs of the American Dental Association (ADA) resolved to develop a Clinical Practice Guideline (CPG) for the non-surgical treatment \citep{Smiley_etal_2015} of CP with SRP, with or without adjuncts, based on a systematic literature review. The panel found 0.5 mm average improvement in CAL with SRP, while the combination of SRP with assorted adjuncts resulted in (average) CAL improvement of 0.2-0.6 mm, over SRP alone. However, comparison among the adjuncts were not conducted. Recently, a systematic network meta-analyses (NMA) of the adjuncts in 74 studies from the CPG revealed none of them to be statistically (significantly) superior to the other \citep{John_etal_2017}. However, the NMA ranked SRP + doxycycline hyclate gel (a local antimicrobial)' as the best non-surgical treatment of CP, compared to SRP alone. Throughout the years, lasers have revolutionized oral care, with reported advantages in minimizing tissue damages, swelling, and bleeding, leading to high patient acceptance. However, it's clinical efficacy, both as an alternative, or adjuvant to SRP \citep{liu1999comparison, zhao2014er} remains inconclusive, with inconsistent evidence derived from underpowered clinical studies \citep{porteous2014adjunctive}; see also the April 2011 statement \citep{perio2011AAP} by the AAP.

Although (randomized) clinical trials (RCT) and subsequent meta-analyses continue to remain the \textit{de facto} in understanding effectiveness of a treatment (or intervention) over others, conducting a successful RCT comes with its own bag of limitations, which includes issues with patient recruitment and retention, escalating costs, educating the wider public, etc. This has led to the development of more \textit{patient-centric} approaches, primarily through adaptive interventions, or dynamic treatment regimes (DTR) \citep{Murphyetal2001,Murphy2003, Robins2004} under the umbrella of precision medicine \citep{garcia2013expanding}, where the focus changed to decision-making for the individual, or subgroups, rather than average-response based traditional RCT treatment comparisons. DTRs involve multistage sequential interventions, according to the patients' evolving characteristics (such as patient's response, or adherence) at each subsequent treatment stage. The treatment types are repeatedly adjusted over time to match an individual's need in order to achieve optimal treatment effect. They are very appealing in managing chronic diseases that require long-term care \citep{Lavorietal2000, MurphyMcKay2003}. Examples of DTRs from various clinical areas include alcoholism \citep{Breslinetal1999}, smoking cessation \citep{Chakraborty_etal_2010}, drug abuse \citep{Brooner2002}, depression \citep{Untzeretal2001}, hypertension \citep{Glasgowetal1989}, etc. However, in the field of oral health and CP, a DTR proposal to mitigate the aforementioned issues related to RCTs seem to be non-existent. For example, in the treatment of CP, one may develop a simple adaptive strategy of continuing the SRP (in the later stages) among the SRP responders-only group (in the first stage), while subjecting the non-responders to `SRP + adjuncts' at later stages.

\begin{figure}[ht]
\begin{center}
\begin{tabular}{cc}
a) \includegraphics[scale=0.5]{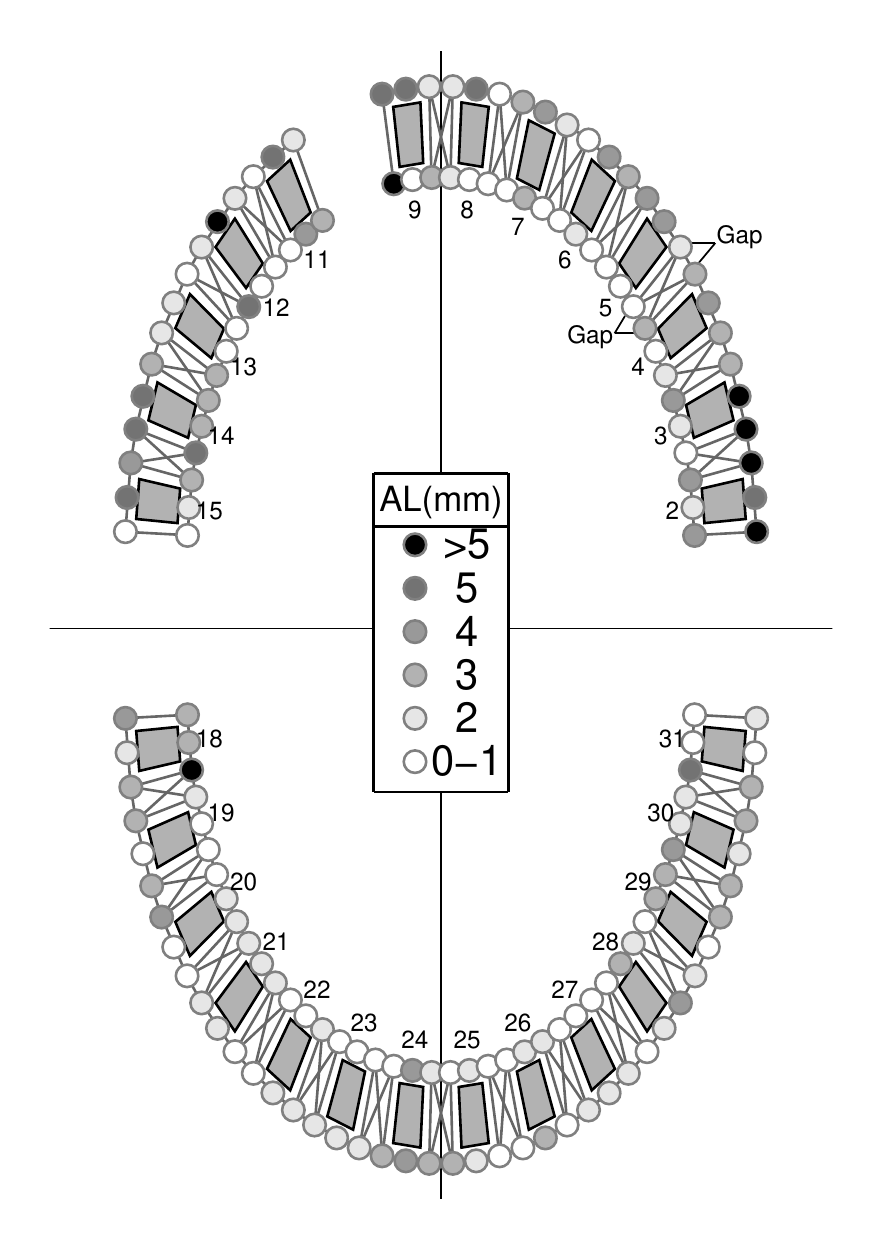} & b) \includegraphics[scale=0.5]{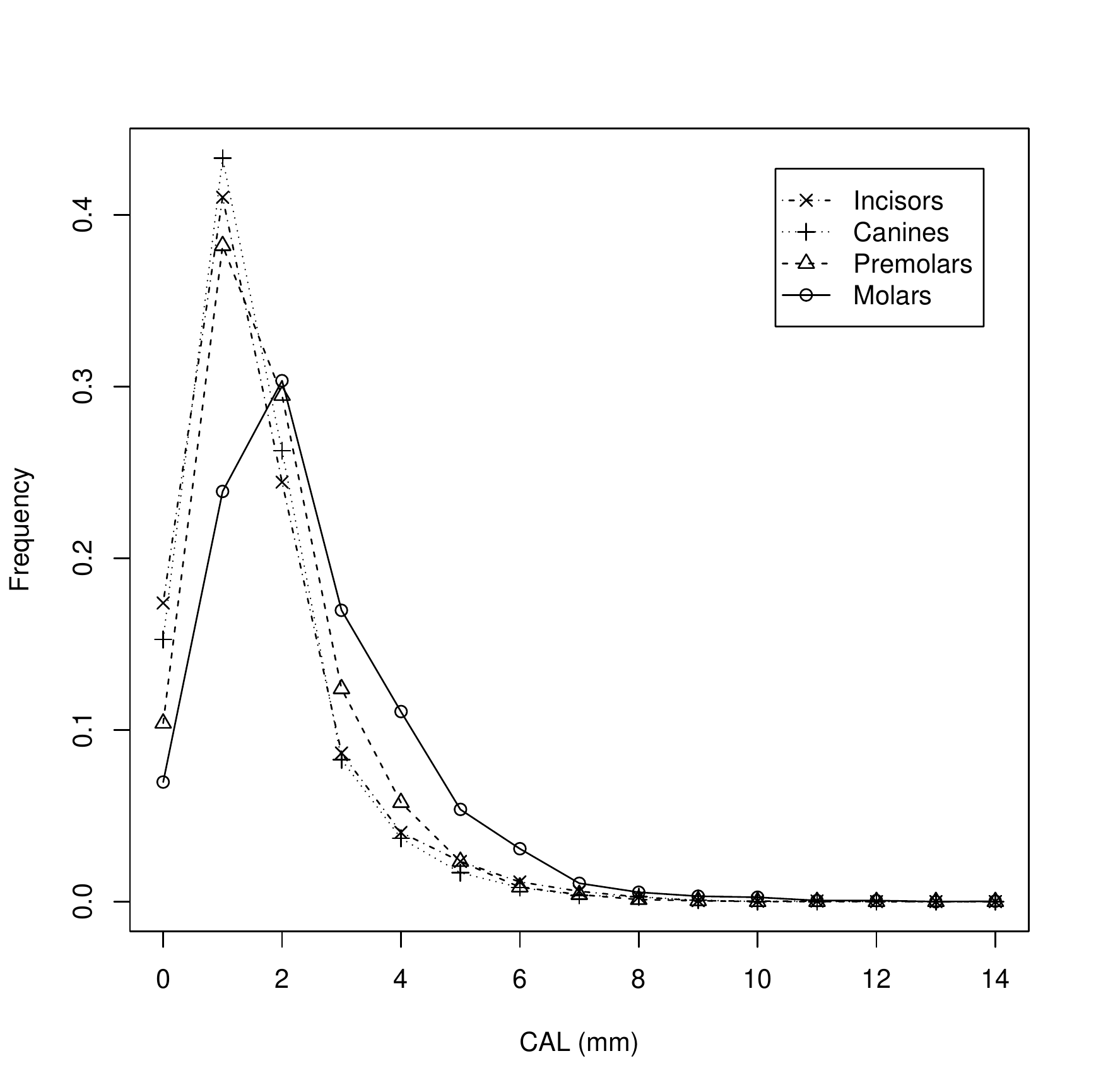}\\
\end{tabular}
\end{center}
\caption{CAL data. Panel (a) shows the observed CAL for a patient with a missing incisor, where the shaded boxes represent teeth, the circles represent sites, and gray lines represent neighbour pairs that connects adjacent sites on the same tooth and sites that share a gap between teeth. ``Gap" in the figure indicates, for example, the four sites in the gap between teeth \# 4 and 5. The tooth numbers are indicated, and excludes the 4 third-molars:  1, 16, 17 32. The vertical and horizontal lines separate the mouth into four quadrants, with the molars (\# 2-3, 14-15, 18-19, 30-31), premolars (\# 4-5, 12-13, 20-21, 28-29), canines (\# 6, 12, 22, 27) and incisors (\# 7-10, 23-26). Panel (b) presents the frequency density plot of the CAL (rounded to the nearest mm) for each tooth type from the GAAD dataset.
}\label{fig:teethCAL}
\end{figure}

A special class of designs, called \textbf{S}equential \textbf{M}ultiple \textbf{A}ssignment \textbf{R}andomized \textbf{T}rial, or SMART designs \citep{Murphy_2005, lei2012smart} are popularly used to study DTRs \citep{Chakraborty_2013}. SMART designs involve randomizing patients to available treatment options at the initial stage, followed by re-randomizing at each subsequent stage of some or all of the patients to treatments available at that stage. These re-randomizations and set of treatment options depend on how well the patient responded to the previous treatments \citep{Ghosh_2016}. Contrary to the standard RCT that takes a `one size fits all' single intervention approach, the SMART advances the RCT by following the same individual over sequential randomizations (i.e., ordered interventions) with the underlying series and order of interventions depicting a real-life setting, that can drastically affect eventual outcomes. Although \cite{Murphy_2005} proposed the general SMART design framework, the treatment was restricted to individual-level outcomes. However, in our motivating clinical discipline of treating CP (and in other behavioral intervention research), interventions are delivered at the group, or cluster (individual) level, while the CAL responses are available at the cluster sub-unit level (teeth) level. Although sample size formulas and SMART design implementations under clustered outcomes setting are available \citep{Ghosh_2016, NeCamp_2017}, they only focus on regimes that do not share an initial treatment. Furthermore, they do not account for other data complications typical to PD studies, such as presence of (i) non-Gaussian (skewed and thick-tailed), (ii) non-randomly missing, and (iii) spatially-referenced CAL responses \citep{Reich_etal_2013}. For example, consider the motivating GAAD data, which recorded the extent of PD in a Type-2 diabetic Gullah-speaking African American population from the coastal South Carolina sea-islands \citep{fernandes2009periodontal}. For illustration, panel (a) in Figure \ref{fig:teethCAL} describe the measurement locations and sample data for a random subject, while panel (b) plots the density histogram of the CAL for the four tooth-types from the GAAD dataset, revealing considerable right-skewness. Furthermore, PD being the major cause of adult tooth-loss, it is likely that patients with higher level of CAL (and CP) exhibit a higher proportion of missing teeth, and hence this missingness mechanism is non-ignorable \citep{Reich_etal_2013}. Also, CP and PD are hypothesized to be spatially-referenced, i.e., proximally located teeth usually have similar disease status than distally located ones. Ignoring the features (i)--(iii) in constructing any SMART design for CP may lead to imprecise estimates of the desired parameters. It is important to note here that the \cite{Ghosh_2016} approach of a clustered SMART design considers traditional clustering (sub-units within a cluster) of Gaussianly distributed continuous responses, and excludes spatial clustering and other features.

In this paper, we set forward to address the aforementioned limitations in developing a list of plausible DTRs for treating CP. We cast this into a two-stage SMART design framework for CP outcomes that exhibit (i)--(iii), and present an analysis plan and sample size calculations for (a) detecting a postulated effect size of a single treatment regime, and (b) detecting a postulated difference between two treatment regimes with or without a shared initial treatment. The tooth-level covariance structure describing spatial-association is modeled by a conditionally autoregressive process \citep{Reich_Bandyopadhyay_2010}. To accommodate possible skewness and tail behavior, the tooth-level CAL responses are assumed to have skew-$t$  \citep{azzalini2003distributions} errors, with the non-randomly missing CAL values imputed via a shared parameter model corresponding to the missingness indicator. The proposed method considers mean comparison for the regimes with or without sharing an initial treatment, where the expected values and corresponding variances or covariance of the effect size of the treatment regimes are derived by the inverse probability weighting (IPW) techniques  \citep{robins1994estimation}, and method of moments.

The rest of the paper is organized as follows. Section \ref{s:dentalSMART-D} introduces eight potential treatment, and the corresponding DTRs that constitute the 2-stage SMART design for CP. Section \ref{s:analysissamplesize} presents the theoretical framework and a sample size calculation method under this SMART design, incorporating the aforementioned features typical to PD data. Section \ref{s:sim} investigates the finite-sample performance of the proposed sample size calculation method using synthetic data generated under various settings. Section \ref{s:demo} demonstrates the implementation of the \texttt{R} function \texttt{SampleSize.SMARTp} for calculating sample sizes, also available at the \texttt{GitHub} link \url{https://github.com/bandyopd/SMARTp}. Finally, the paper ends with a discussion in Section \ref{s:disc}. Supplementary Material, consisting of detailed derivations are relegated to the Appendix.


\section{A SMART design for the DTRs} \label{s:dentalSMART-D}

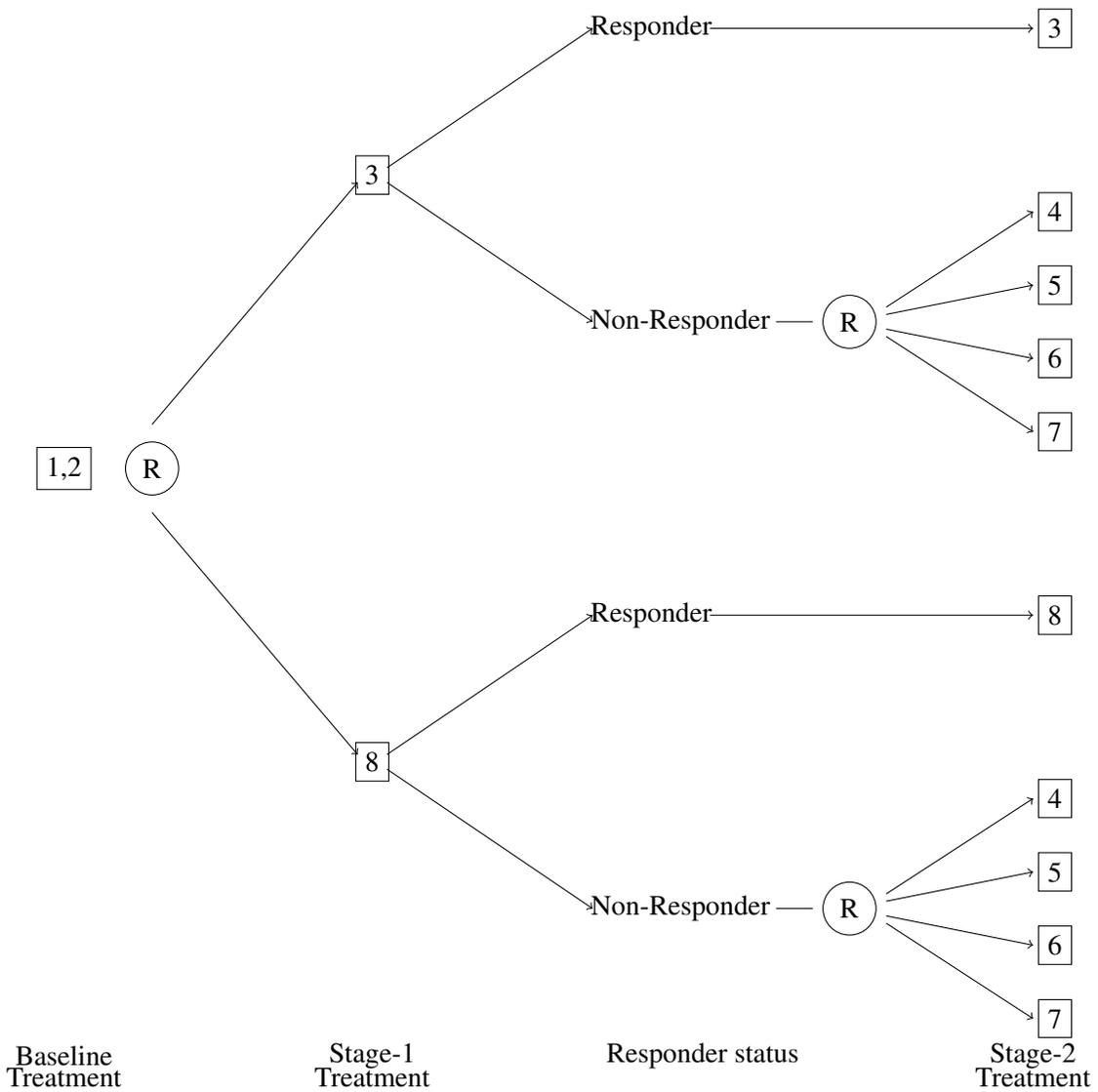
\begin{figure}
\centering
\begin{tikzpicture}
\boldmath


\node at (0,0)[circle,draw] {R};
\node at (-1.2,0)[rectangle,draw] {1,2};
\node at (3,4)[rectangle,draw] {3};
\node at (3,-4)[rectangle,draw] {8};

\draw [->] (0,0.6) -- (2.8,3.9);
\draw [->] (0,-0.6) -- (2.8,-3.9);

\draw [->] (3.2,4.1) -- (6,6);
\node at (6.8,6) {{Responder}};
\draw [->] (7.6,6) -- (12,6);
\node at (12.3,6)[rectangle,draw] {3};

\draw [->] (3.2,3.9)   -- (6,2);
\node at (7.2,2) {Non-Responder};
\draw [-] (8.5,2) -- (9, 2);
\node at (9.5,2)[circle,draw] {R};

\draw [->] (10,2.2) -- (12,3.5);
\node at (12.3, 3.5)[rectangle,draw] {4};
\draw [->] (10,2.1) -- (12,2.5);
\node at (12.3, 2.5)[rectangle,draw] {5};
\draw [->] (10,1.9) -- (12,1.5);
\node at (12.3, 1.5)[rectangle,draw] {6};
\draw [->] (10,1.8) -- (12,0.5);
\node at (12.3, 0.5)[rectangle,draw] {7};

\draw [->] (3.2,-3.9) -- (6,-2);
\node at (6.8,-2) {{Responder}};
\draw [->] (7.6,-2) -- (12,-2);
\node at (12.3,-2)[rectangle,draw] {8};

\draw [->] (3.2,-4.1) -- (6,-6);
\node at (7.2,-6) {Non-Responder};
\draw [-] (8.5,-6) -- (9,-6);
\node at (9.5,-6)[circle,draw] {R};

\draw [->] (10,-5.8) -- (12,-4.5);
\node at (12.3, -4.5)[rectangle,draw] {4};
\draw [->] (10,-5.9) -- (12,-5.5);
\node at (12.3, -5.5)[rectangle,draw] {5};
\draw [->] (10,-6.1) -- (12,-6.5);
\node at (12.3, -6.5)[rectangle,draw] {6};
\draw [->] (10,-6.2) -- (12,-7.5);
\node at (12.3, -7.5)[rectangle,draw] {7};

\node at (-1.2,-8) {Baseline} ;
\node at (3,-8) {Stage-1} ;
\node at (7.5,-8) {Responder status} ;
\node at (12,-8) {Stage-2} ;

\node at (-1.2,-8.3) {Treatment} ;
\node at (3,-8.3) {Treatment} ;
\node at (12,-8.3) {Treatment} ;
\end{tikzpicture}
\caption{A SMART design diagram for developing DTRs treating chronic periodontitis. R=randomization, 1=oral hygiene instruction, 2=education on risk reduction, 3=scaling and root planing (SRP), 4=SRP with local antimicrobial therapy, 5=SRP with systemic antimicrobial therapy, 6=SRP with photodynamic therapy, 7=SRP with systemic sub-antimicrobial-dose doxycycline (SDD), 8=laser therapy.} \label{fig:smart_dental}
\end{figure}

In this section, we propose dynamic treatment regimes (DTRs) for treating CP, which are studied via a SMART design. A list of possible treatments consist of the treatment initiation steps: (1) Oral hygiene instruction, and (2) Education on risk reduction. This is followed by (3) SRP, or more advanced non-surgical treatments that combine SRP with adjunctive therapy, as summarized in the systematic review of \cite{Smiley_etal_2015},  such as (4) SRP with local antimicrobial therapy, (5) SRP with systemic antimicrobial therapy, (6) SRP with photodynamic therapy, which uses lasers, but only to activate an antimicrobial agent), (7) SRP with systemic subantimicrobial-dose doxycycline (SDD), and finally, (8) Laser. The corresponding SMART design for developing DTRs is presented in Figure \ref{fig:smart_dental}. Note that the number of potential DTRs are not limited to Figure \ref{fig:smart_dental}. More details are presented in Section \ref{s:disc}.

\vskip.1in

Oral hygiene is primarily used for prevention and initial therapy, especially during early stage of periodontitis. At the beginning of the proposed trial, each participant has to attend the treatment initiation steps (1) and (2) before any randomization. Note that while SRP is the accepted gold-standard, the role of laser therapy, though advantageous in targeting the diseased area precisely and accurately, still remains controversial as a standard of care. In this paper, we develop our SMART design, with a primary focus on comparing the DTRs starting with either SRP (\# 3), or Laser therapy (\# 8).  At the initial stage, each participant is randomly allocated to either treatment 3 or 8. We propose a DTR that matches an patient's's need in achieving similar outcome as SRP with adjuncts, though at a lower cost. Each possible treatment regime can have more than one path of treatment according to each patient's evolving response. The patients who respond to the initial treatment continue the same treatment at the 2nd stage of the trial. The patients who do not respond to treatment 3 are randomly allocated to one of the treatments 4--7 in the 2nd stage. Similarly, for patients allocated to the laser arm (treatment 8), the non-responders will also have the provision of being randomly allocated to one of 4--7 in the 2nd stage. The randomization probabilities calculated at both the initial and final stages of our SMART design is presented in Section \ref{samplesize}. The primary final outcome measure is the recorded and rounded tooth-level CAL. The possible paths are listed below, i.e.
\begin{itemize}
\item Path 1: `1', `2', `3', `3'
\item Path 2: `1', `2', `3', `4'
\item Path 3: `1', `2', `3', `5'
\item Path 4: `1', `2', `3', `6'
\item Path 5: `1', `2', `3', `7'
\item Path 6: `1', `2', `8', `8'
\item Path 7: `1', `2', `8', `4'
\item Path 8: `1', `2', `8', `5'
\item Path 9: `1', `2', `8', `6'
\item Path 10: `1', `2', `8', `7'
\end{itemize}

This leads to eight different DTRs ($d_1$-$d_8$) that are embedded within the two-stage SMART design, i.e.

\vskip.1in\noindent Regime 1 $(d_1)$: (`3',`3'$^R$,`4'$^{NR}$)
\vskip.1in\noindent Regime 2 $(d_2)$: (`3',`3'$^R$,`5'$^{NR}$)
\vskip.1in\noindent Regime 3 $(d_3)$: (`3',`3'$^R$,`6'$^{NR}$)
\vskip.1in\noindent Regime 4 $(d_4)$: (`3',`3'$^R$,`7'$^{NR}$)
\vskip.1in\noindent Regime 5 $(d_5)$: (`8',`8'$^R$,`4'$^{NR}$)
\vskip.1in\noindent Regime 6 $(d_6)$: (`8',`8'$^R$,`5'$^{NR}$)
\vskip.1in\noindent Regime 7 $(d_7)$: (`8',`8'$^R$,`6'$^{NR}$)
\vskip.1in\noindent Regime 8 $(d_8)$: (`8',`8'$^R$,`7'$^{NR}$)
\\

Here,  Regime 1 can be explained as following treatments 1 and 2, a patient undergoes treatment 3 (considered as treatment at initial stage). If that patient responds (R) to the initial treatment, he or she continues with treatment 3 at 2nd stage, while a non-responder (NR) will receive treatment 4 at the 2nd stage. The other regimes can be explained similarly. There are a number of advantages \citep{Chakraborty_2013} of SMART designs over a series of single-stage trials to develop an optimal DRT. First, the single-stage trials may fail to detect the delayed effect. For example, the patients who respond to laser therapy may achieve better outcomes than those who respond to SRP initially, however, the effectiveness of the SRP may be realized in the later stages (here, 2nd stage) when possible adverse events may occur due to laser therapy. Second, the single-stage trials may also fail to detect the diagnostic effect. Based on the patients' treatment outcome at initial stage, e.g., laser therapy, the SMART design could allocate the right treatment at final stage depending on participants' response, e.g., the non-responders of laser therapy receive SRP with one of the adjuncts, e.g, systematic anti-microbial at final stage. Third, single-stage trials may result in possible cohort effect. The non-responding patients who are treated by SRP at initial stage may drop out of single-stage trials, which could incur biased estimation of treatment effect. Under the proposed SMART design, participants expect to receive better treatments, i.e. SRP with adjuncts, at the next stage, if SRP is not effective initially.

\section{SMART Design: Model, Hypothesis Testing and Sample Size Calculations}\label{s:analysissamplesize}

In this section, we propose the theoretical framework and a novel sample size formula for our SMART design.

\subsection{Statistical Model}\label{s:analysis}

We start with introducing some notations. Let $A_{i1}$ denote the treatment for patient $i$ at the initial stage (i.e. `3' or `8'); $R_{i}(A_{i1})$ denote the proximal response after initial treatment $A_{i1}$, i.e $R_{i}(\cdot)=1$ if the $i^{\text{th}}$ patient is a responder and $R_{i}(\cdot)=0$ otherwise; $A_{i2}(A_{i1}, R_{i}(A_{i1}))$ denote the treatment at final stage based on initial (first) stage treatment and response; $Y_{it}$ denote the final outcome measure, i.e. \textit{change in mean CAL} for the $t^{\text{th}}$ teeth of patient $i$; $M_{it}$ denotes the missingness indicator of the $t^{\text{th}}$ teeth of patient $i$, i.e. $M_{it}=1$ if missing, or $0$ otherwise.
Thus, the observed data trajectory for patient $i$ can be described as $\boldsymbol O_{i}$=($A_{i1}$, $R_{i}(A_{i1})$, $A_{i2}(A_{i1}, R_{i}(A_{i1}))$, $Y_{i,1}$,$\ldots$,$Y_{i,28}$, $M_{i,1}$,$\ldots$,$M_{i,28}$). Note that we have $N$ patient in the sample, and each patient has a maximum of $28$ teeth (if no tooth is missing). Thus, the (overall) outcome measure
for patient $i$ is $\bar{Y}_i=\sum_{t=1}^{28}Y_{it}(1-M_{it})/\sum_{t=1}^{28}(1-M_{it})$, which is the mean of CAL of the \textit{available teeth} for patient $i$. Hence the proportion of the available teeth for patient `$i$' is $\hat{p}_i=\sum_{t=1}^{28}(1-M_{it})/28$. The regression model for $\boldsymbol Y$ is given by:


\begin{equation}\label{regY}
Y_{it}=\mu_i + Q_{it}+\epsilon_{it1},\\
\end{equation}
for $i=1,\ldots,N$ and $t=1,\ldots,28$, where $\mu_i$ = $\beta_{0}+\beta_{1}A_{i13}+\beta_{2}A_{i13}R_i+\beta_{3}R_i+\beta_{4}A_{i13}A_{i24}(1-R_i)+\beta_{5}A_{i13}A_{i25}(1-R_i)+\beta_{6}A_{i13}A_{i26}(1-R_i)$.
Here, $A_{i13}$ is an indicator of treatment `3' at initial stage for patient $i$, $A_{i24}$ is an indicator of treatment `4' at final stage for patient $i$, and $\epsilon_{it1}$ is the (random) error term distributed as a skew-normal (SN(, or skew-$t$ (ST) density \citep{Azzalini_Capitanio_2003}, i.e., $\epsilon_{it1} \sim ST(0, \sigma_{1}^{2}, \lambda, \nu)$, with location parameter $0$, scale parameter $\sigma_{1}$, skewness parameter $\lambda$, and degrees of freedom $\nu$ that measure the kurtosis. Note, the distribution of $\epsilon_{it1}$ is normal if $\lambda=0$ and $\nu=\infty$, skew-normal if $\lambda\neq 0$ and $\nu=\infty$, $t$ if $\lambda=0$ and $\nu<\infty$, and skew-$t$, if $\lambda\neq 0$ and $\nu<\infty$. Expressions of the mean, variance, skewness $\gamma_1$ and kurtosis $\gamma_2$ for both SN and ST distributions are presented in Appendices \ref{SN} and \ref{ST} respectively. Following \cite{Reich_etal_2013}, we assume the latent vector $\boldsymbol Q_i$=$(Q_{i1},\ldots,Q_{i28})^{\top}$ follows a multivariate normal distribution, with mean vector $\boldsymbol 0_{28\times 1}$ and covariance matrix $\boldsymbol\Sigma_{28\times 28}$ with a conditional autoregressive (CAR) structure, i.e. $\boldsymbol\Sigma_{28\times 28}=\tau^2(\boldsymbol C_{28\times 28}-\rho \boldsymbol D_{28\times 28})^{-1}$. Here, $\tau^2>0$ and $\rho\in[0,1]$ are the parameters controlling the magnitude of variation, and degree of spatial association, respectively. For matrix $\boldsymbol D$, the elements $D_{tt^{\prime}}$ are ones if locations $t$ and $t^{\prime}$ are adjacent, and zeroes otherwise. The matrix $\boldsymbol C$ is diagonal with diagonal elements $C_{tt}=\sum_{t^{\prime}}D_{tt^{\prime}}$.

\begin{sloppypar}
Next, under the assumption of non-randomly missing teeth (locations of missing teeth are not random, but rather related to the CP health in that region of the mouth), we propose a probit regression model for the missing teeth indicator as a function of the underlying (spatial) latent term $\boldsymbol Q_i$. Define $M_{it} = I(M_{it0} > 0)$, where $M_{it0}$ is a (latent) continuous variable, modeled as:

\begin{equation}\label{regM}
M_{it0} = a_{0}+ b_{0}Q_{it}+\epsilon_{it0}, \\
\end{equation}
where $\epsilon_{it0}\stackrel{i.i.d}{\sim} N(0, \sigma_{0}^{2})$. For sake of identifiability, we choose $\sigma_{0}^{2} = 1$. Here, under the popular shared-parameter framework \citep{vonesh2006shared}, $\boldsymbol Q$ facilitates sharing of information between $\boldsymbol Y$ and $\boldsymbol M$ for modeling non-randomly missing data.  The parameters $a_0$, $b_0$, and the estimates $\boldsymbol Q_{i}$ and $\epsilon_{it0}$ determine the proportion of available tooth $p_i=E(\hat{p}_i)$, which can be estimated using either stochastic or deterministic method (see Appendix C). The parameter $b_0$ controls the association between $\boldsymbol Y$ and $\boldsymbol M$, e.g., $b_0=0$ indicates no association. For example, the Pearson correlation coefficient between $Y_{it}$ and $M_{it0}$ (from (\ref{regM})) is $c_{it}$ = $b_0\text{var}(Q_{it})/\sqrt{ ( \text{var}(Q_{it}) + \text{var}(\epsilon_{it1}) ) ( b_0^2\text{var}(Q_{it}) + \text{var}(\epsilon_{it0}) ) }$, see Appendix C for the derivation. For power analysis, one may choose the distributions of $\boldsymbol Q_{i}$, $\epsilon_{it1}$ and $\epsilon_{it0}$ from the literature, e.g., \cite{Reich_Bandyopadhyay_2010}. Define $c_i$=$\sum_t c_{it}/28$. The clinician may suggest values for $\mu_i$, $p_i$ and $c_i$, and the corresponding estimates of $a_0$ and $b_0$ can be obtained by solving aa set of simultaneous equations involving $p_i$ and $c_i$.

Next, we derive the expected value and variance for the sample mean of a DTR, using $d_1$ as an example, based on the IPW principle. IPW techniques have been successfully applied for estimating regression coefficients \citep{Robins_etal_1994}, and population mean \citep{Cao_etal_2009}, in the context of incomplete data. For DTRs under SMART designs, most likely, we are unable to sample data directly from a particular regime. For example, responders of SRP can be classified as either regimes 1-4. Hence, a method of moments estimate of the sample mean for regime 1 is given by:

\begin{equation}\label{Eyd1}
\bar{Y}^{d_1} = E_{d_1}(\bar{Y}_{i}) = E\left( W_{i}^{d_1}\bar{Y}_{i} \right),
\end{equation}

where
\begin{equation}\label{Wid1}
W_{i}^{d_1}=\dfrac{ I( A_{i1}=a_{1i}^{d_1}, A_{i2}=[a_{2i}^{d_1 R}]^{R_i}[a_{2i}^{d_1 NR}]^{1-R_i} ) }{
                    \pi_{1i}^{d_1} [\pi_{2i}^{d_1 R}]^{R_i}[\pi_{2i}^{d_1 NR}]^{1-R_i} }.
\end{equation}

\noindent where, $\bar{Y}_{i}$: the final outcome measure of CAL change for patient $i$; $I(\cdot)$: indicator function; $R_i$: binary response indicator for treatment `3' at initial stage; $a_{1i}^{d_1}$: the regime 1 ($d_1$) treatment at initial stage for patient $i$, e.g., `3'; $a_{2i}^{d_1 R}$: regime 1 treatment at final stage if participant $i$ is a responder (i.e. $R_i=1$), e.g., `3'; $a_{2i}^{d_1 NR}$: regime 1 treatment at final stage if patient $i$ is a non-responder (i.e. $R_i=0$), e.g., `4'; $\pi_{1i}^{d_1}$: probability of treatment allocation of regime 1 at initial stage for patient $i$; $\pi_{2i}^{d_1 R}$: probability of treatment allocation of regime 1 at final (2nd) stage if patient $i$ is a responder, i.e. 1; $\pi_{2i}^{d_1 NR}$: probability of treatment allocation of regime 1 at final stage, if patient $i$ is a non-responder, i.e. 1/4.
\end{sloppypar}

To maximize power, we estimate $\pi_{1i}^{d_1}$ as in \citep{Murphy_2005} to have equal sample sizes across all possible regimes. We set

\begin{equation}\label{pi1}
\pi_{1i}^{d_1}=\dfrac{( 1\cdot\gamma^{d_1}+ \dfrac{1}{4}\cdot (1-\gamma^{d_1}) )^{-1}}{
                      ( 1\cdot\gamma^{d_1}+ \dfrac{1}{4}\cdot (1-\gamma^{d_1}) )^{-1} +
                      ( 1\cdot\gamma^{d_5}+ 1\cdot (1-\gamma^{d_5}) )^{-1} },
\end{equation}

where $\gamma^{d_1}$ denotes the response rate for regime 1 at initial stage. If $\gamma^{d_1}$ and $\gamma^{d_5}$ are not known, we set

\begin{equation}
\pi_{1i}^{d_1}=\dfrac{ \text{max}( 1^{-1}, \dfrac{1}{4}^{-1} ) }{
                       \text{max}( 1^{-1}, \dfrac{1}{4}^{-1} ) +
                       \text{max}( 1^{-1}, 1^{-1} ) }.
\end{equation}
Alternatively, we set $\pi_{1i}^{d_1}=1/2$, if equal probability of treatment allocation at initial stage is required.

The mean and variance of $\bar{Y}^{d_1}$ are derived below. We have

\begin{equation}\label{Eybard1}
E(\bar{Y}^{d_1})=E( \dfrac{ I( A_{i1}=a_{1i}^{d_1}, A_{i2}=[a_{2i}^{d_1 R}]^{R_i}[a_{2i}^{d_1 NR}]^{1-R_i} ) }{ \pi_{1i}^{d_1} [\pi_{2i}^{d_1 R}]^{R_i}[\pi_{2i}^{d_1 NR}]^{1-R_i} }\bar{Y}_{i} ),
\end{equation}

\begin{equation}\label{varybard1}
\text{var}(\bar{Y}^{d_1})=\dfrac{1}{N}\text{var}( \dfrac{ I( A_{i1}=a_{1i}^{d_1}, A_{i2}=[a_{2i}^{d_1 R}]^{R_i}[a_{2i}^{d_1 NR}]^{1-R_i} ) }{ \pi_{1i}^{d_1} [\pi_{2i}^{d_1 R}]^{R_i}[\pi_{2i}^{d_1 NR}]^{1-R_i} }\bar{Y}_{i} ).
\end{equation}

In terms of $\boldsymbol\pi$, $\boldsymbol\gamma$, $\boldsymbol\mu$ and $\boldsymbol\sigma$, (\ref{Eybard1}) and (\ref{varybard1}) can be expressed alternatively as

\begin{align*}
E(\bar{Y}^{d_1})=\gamma^{d_1}\mu_{d_{1}R}+(1-\gamma^{d_1})\mu_{d_{1}NR}=\mu_{d_{1}}
\end{align*}

and

\begin{align*}
&V(\bar{Y}^{d_1}) \\
=\dfrac{1}{N} & \{ \dfrac{\gamma^{d_1}}{\pi_{1i}^{d_1} [\pi_{2i}^{d_1 R}]}( \sigma_{d_{1} R}^{2} + ( 1-\pi_{1i}^{d_1} [\pi_{2i}^{d_1 R}] )\mu_{d_{1} R}^{2} ) +  \\
&\dfrac{1-\gamma^{d_1}}{\pi_{1i}^{d_1} [\pi_{2i}^{d_1 NR}]}( \sigma_{d_{1}NR}^{2} + ( 1-\pi_{1i}^{d_1} [\pi_{2i}^{d_1 NR}] )\mu_{d_{1}NR}^{2} )+ \\
&\gamma^{d_1}(1-\gamma^{d_1})(\mu_{d_{1}R}- \mu_{d_{1}NR})^{2} \},
\end{align*}

\noindent where $\sigma_{d_{1}R}^{2}$: the variance of $\bar{Y}_i$ from $d_1$ and $R_i=1$; $\mu_{d_{1}R}$: the mean of $\bar{Y}_i$ from $d_1$ and $R_i=1$. Detailed expressions of both $\sigma_{d_{1}R}^{2}$ and $\mu_{d_{1}R}$ appear in Appendix \ref{SamFormDerApp}. However, we compute them using Monte Carlo method. Likewise, both $E(\bar{Y^{d_3}})$ and $V(\bar{Y}^{d_3})$ can be derived; see Appendix \ref{SamFormDerApp}.


\subsection{Hypothesis Testing}\label{testing}

Our SMART design allows the following three important hypothesis tests:

\begin{enumerate}
\item[1.]Detecting a single DTR effect on CAL, e.g., $H_0:~\mu_{d_1}=0$ vs $H_1:~\mu_{d_1}=\delta_{d_1}\neq 0$;
\item[2.]Comparing two DTRs that share an initial treatment, i.e. DTRs with SRP and adjuncts, e.g., $H_0:~\mu_{d_1}-\mu_{d_3}=0$ vs $H_1:~\mu_{d_1}-\mu_{d_3}=\delta_{d_1-d_3}\neq 0$;
\item[3.]Comparing two treatment DTRs that do not share an initial treatment, i. e. the DTRs initialized by SRP and laser, e.g. $H_0:~\mu_{d_1}-\mu_{d_5}=0$ vs $H_1:~\mu_{d_1}-\mu_{d_5}=\delta_{d_1-d_5}\neq 0$.
\end{enumerate}

Hypothesis 1 can be used to test whether the improvement in CAL in the proposed DTR is better than SRP (e.g., $\geq 0.5$mm), or not worse than SRP with adjuncts (e.g., 0.7--1.1 mm), based on the systematic review results of Smiley \textit{et al.} \cite{Smiley_etal_2015}. Since the network meta-analyses by John \textit{et al.} \cite{John_etal_2017} found no significant evidence of CAL improvement among adjuncts, Hypothesis 2 can be used to test if indeed there are statistically significant differences between the DTRs of SRP and `SRP + adjuncts'. The treatment effect of laser therapy is still under investigation; we can use Hypothesis 3 to test if there is a statistically significant difference between the DTRs initialized by SRP and laser.

Consider Hypothesis 2. The Expectation and variance of regimes difference can be expressed respectively as
\begin{align}\label{delta}
E(\bar{Y}^{d_1}-\bar{Y}^{d_3})=\mu_{d_1}-\mu_{d_3}=\delta_{d_1-d_3}
\end{align}
and
\begin{align}\label{sigmaE}
V(\bar{Y}^{d_1}-\bar{Y}^{d_3})=V(\bar{Y}^{d_1})+V(\bar{Y}^{d_3})-2COV(\bar{Y}^{d_1}, \bar{Y}^{d_3})=\dfrac{1}{N}2\sigma_{d_1-d_3}^{2}.
\end{align}

Note, both $\delta$ and $\sigma^2$ in equations (\ref{delta}) and (\ref{sigmaE}) respectively are functions of the parameter vector $\boldsymbol\Omega_{d_1-d_3}$=($\mu$, $\tau$, $\rho$, $\lambda$, $\nu$, $\sigma_1^2$, $\sigma_0^2$, $a_0$, $b_0$, $\gamma^{d_1}$, $\pi_{1}^{d_1}$, $\pi_{2}^{d_1 R}$, $\pi_{2}^{d_1 NR}$, $\gamma^{d_3}$, $\pi_{1}^{d_3}$, $\pi_{2}^{d_3 R}$, $\pi_{2}^{d_3 NR}$). Note, $\mu$, $\tau$, $\rho$, $\lambda$, $\nu$, $\sigma_1^2$, $\sigma_0^2$, $a_0$ and $b_0$ are defined in equations (\ref{regY}) and (\ref{regM}), while parameters $\gamma$'s and $\pi$'s are defined in equations (\ref{Eyd1}) to (\ref{pi1}). The covariance between $\bar{Y}^{d_1}$ and $\bar{Y}^{d_3}$ is

\begin{align*}
&\text{COV}\left(\bar{Y}^{d_1}, \bar{Y}^{d_3}\right)\\
=&\dfrac{1}{N}\{\dfrac{\gamma^{d_1}}{\pi_{1i}^{d_1} \pi_{2i}^{d_1 R}}(\sigma_{d_1 R}^{2} + \mu_{d_1 R}^{2} ) -\gamma^{d_1}\gamma^{d_3} \mu_{d_1 R}\mu_{d_3 R}\\
&-\gamma^{d_1}(1-\gamma^{d_3})\mu_{d_1 R}\mu_{d_3 NR}\\
&-\gamma^{d_3}(1-\gamma^{d_1})\mu_{d_1 NR}\mu_{d_3 R}\\
&-(1-\gamma^{d_1})(1-\gamma^{d_3})\mu_{d_1 NR}\mu_{d_3 NR}\}.\\
\end{align*}

Note that $\gamma^{d_1}=\gamma^{d_3}$, $\mu_{d_1 R}=\mu_{d_3 R}$ and $\sigma_{d_1 R}^{2}=\sigma_{d_3 R}^{2}$, since the responders from treatment `3' are consistent with both the treatment regimes 1 and 3. The derivations of both $E(\bar{Y}^{d_1}-\bar{Y}^{d_3})$ and $V(\bar{Y}^{d_1}-\bar{Y}^{d_3})$ can be found in Appendix \ref{SamFormDerApp}. We now present a theoretical result below; see it's proof in Appendix \ref{ThePro}.

\begin{thrm} \label{theo1}
The IPW and MOM estimator $\hat{\delta}_{d_1-d_3}$ is a consistent estimator of $\delta_{d_1-d_3}$. Under moment conditions and following assumptions below, we have $\sqrt{N}(\hat{\delta}_{d_1-d_3}-\delta_{(d_1-d_3)0})\rightarrow N(0, 2\sigma_{d_1-d_3}^2)$.
\end{thrm}

\noindent \textbf{Assumptions:}
\begin{enumerate}
\item[1.] Random vectors ($\bar{Y}_i$, $W^{d_1}_{i}$, $W^{d_3}_i$), $0\leq i\leq N$ are independent and identically distributed, and distribution of $\bar{Y}_i$ is independent of $W^{d_1}_{i}$ and $W^{d_3}_{i}$, where $W^{d_1}_{i}$ is defined by equation (\ref{Wid1});
\item[2.] $E(W^{d_1}_{i}\bar{Y}_{i} - \mu_{d_1}) = 0$, only when $\mu_{d_1}=\mu_{d_10}$ and $E(W^{d_3}_{i}\bar{Y}_{i} - \mu_{d_3}) = 0$ only when $\mu_{d_3}=\mu_{d_30}$. Hence, $E(\hat{\delta}_{d_1-d_3}-\delta_{d_1-d_3})=0$ only for $\delta_{d_1-d_3}=\delta_{(d_1-d_3)0}$=$\mu_{d_10}-\mu_{d_30}$;
\item[3.] The possible sets for regime means and effect size $\mu_{d_1}$, $\mu_{d_3}$, $\delta_{d_1-d_3}\in\Theta$ are compact;
\item[4.] $\hat{\delta}_{d_1-d_3}-\delta_{d_1-d_3}$ is continuous at each $\delta$, with probability one;
\item[5.] $E_{\text{sup}_{\delta_{d_1-d_3}\in\Theta}}(\hat{\delta}_{d_1-d_3}-\delta_{d_1-d_3})<\infty$.
\end{enumerate}

Though the regimes 1 and 5 do not share any initial treatments, the covariance between the sample mean of these two regimes can be derived in the similar way as $\text{COV}\left(\bar{Y}^{d_1}, \bar{Y}^{d_3}\right)$. The mathematical formula for $\text{COV}\left(\bar{Y}^{d_1}, \bar{Y}^{d_5}\right)$ is given in Appendix \ref{SamFormDerApp}.


Before deriving the sample size formula, we present the test statistics for the corresponding hypotheses below. For example, for $H_0:~\mu_{d_1}-\mu_{d_3}=0$ vs alternative $H_{1}:~\mu_{d_1}-\mu_{d_3}=\delta_{d_1-d_3}\neq 0$ (hypothesis 2), we use the univariate Wald statistics $Z = \delta_{d_1-d_3}/\sqrt{2\sigma_{d_1-d_3}^2/N}$, where $\delta_{d_1-d_3}$ is the effect size and $\sigma_{d_1-d_3}^2$ is given in (\ref{sigmaE}). In large samples, $Z$ follows a standard normal distribution if $H_0$ is true. Hence, at $\alpha$ level of significance, we reject $H_0$ if $\mid Z \mid>z_{\alpha/2}$, where $z_{\alpha/2}$ is the upper $\alpha/2$ quantile of a standard normal distribution. In a similar way, the test statistic for Hypothesis 1 ($H_0:~\mu_{d_1}=0$ vs $H_1:~\mu_{d_1}=\delta_{d_1}\neq 0$) and 3 ($H_0:~\mu_{d_1}-\mu_{d_5}=0$ vs $H_1:~\mu_{d_1}-\mu_{d_5}=\delta_{d_1-d_5}\neq 0$) are $Z=\delta_{d_1}/\sqrt{2\sigma_{d_1}^2/N}$ and $Z=\delta_{d_1-d_5}/\sqrt{2\sigma_{d_1-d_5}^2/N}$ respectively, where both follow standard normal distribution if $H_0$ is true.

\subsection{Sample size calculation}\label{samplesize}




The calculated sample size is possible to detect the effect size of either a single regime or the difference between two regimes. The proposed sample size formulas for Hypothesis tests 1-3 under our SMART design are given by

\begin{equation}\label{smartNclustH1}
N=2(z_{\alpha/2}-z_{1-\beta})^2\dfrac{\sigma_{d_1}^{2}}{\delta_{d_1}^2},
\end{equation}

\begin{equation}\label{smartNclustH2}
N=2(z_{\alpha/2}-z_{1-\beta})^2\dfrac{\sigma_{d_1-d_3}^{2}}{\delta_{d_1-d_3}^2}
\end{equation}

and

\begin{equation}\label{smartNclustH3}
N=2(z_{\alpha/2}-z_{1-\beta})^2\dfrac{\sigma_{d_1-d_5}^{2}}{\delta_{d_1-d_5}^2}
\end{equation}

respectively, where $\sigma_{d_1-d_3}^{2}$ is defined by (\ref{sigmaE}), and in the similar way, both $\sigma_{d_1}^{2}$ and $\sigma_{d_1-d_5}^{2}$ can also be defined; $\alpha = $Pr(Type one error), $\beta = $Pr(Type two error)= 1 - Power, $\text{Pr}(z>z_{\alpha/2})=\alpha/2$ and $\text{Pr}(z>z_{1-\beta})=1-\beta$, the effect size $\delta_{d_1-d_3}=\mu_{d_1}-\mu_{d_3}$. Therefore, we define the standardized effect size by $\delta_{d_1-d_3}^{\ast}=\delta_{d_1-d_3}/\sigma_{d_1-d_3}$. Note, our calculations advance the previous ones for SMART designs in clustered data \citep{Ghosh_2016,NeCamp_2017} by including non-Gaussianity, spatial association, and non-random missingness features, typical for periodontal responses, in addition to considering comparisons between regimes that shares the same initial treatment. Also, the patients (or clusters) are randomly allocated with equal probability for each regime, which requires smaller sample size than allocation with equal treatment probability at each stage.



\section{Simulation studies}\label{s:sim}

We now present simulation studies to investigate the finite-sample performance of the proposed sample size formulas (\ref{smartNclustH1}) to (\ref{smartNclustH3}) in terms of computing Monte Carlo power estimates based on $5000$ simulated data sets, given the type-II error rate $\beta=0.2$, or nominal power of $80\%$ and type-I error rate $\alpha=0.05$. We also compare the theoretical and Monte Carlo mean and variance of the estimated effect sizes for the DTRs.

The Monte Carlo data generation steps are given below. These include generating the random variables $A_{i1}$, $R_{i}(A_{i1})$ and $A_{i2}(A_{i1}, R_{i}(A_{i1}))$, $M_{it}$ and $Y_{it}$ for each patient $i$, $i=1,\ldots,N$.

\begin{enumerate}
\item[Step 1] The initial treatment $A_{i1}$ is assigned randomly to either `3' or `8', with probability $\pi_{1i}^{d_1}$ and $1-\pi_{1i}^{d_1}$ respectively.

\item[Step 2] The response variable $R_{i}(A_{i1})$ is generated from Bernoulli($\gamma^{d_1}$) if $A_{i1}$=`3', or Bernoulli($\gamma^{d_5}$) if $A_{i1}$=`8', where $\gamma^{d_1}=$ 0.25 or 0.5 and $\gamma^{d_5}=0.5$.

\item[Step 3] The final treatment $A_{i2}(A_{i1}=\text{`3'}, R_{i}(A_{i1}=\text{`3'})=1)$ is assigned to `3' with probability of 1, and $A_{i2}(A_{i1}=\text{`3'}, R_{i}(A_{i1}=\text{`3'})=0)$ is randomly assigned to`4', `5', `6' or `7' with probability of 1/4, while $A_{i2}(A_{i1}=\text{`8'}, R_{i}(A_{i1}=\text{'8'})=1)$ is assigned to `8' with probability 1 and $A_{i2}(A_{i1}=\text{`8'}, R_{i}(A_{i1}=\text{'8'})=0)$ is assigned to `4', `5', `6' or `7' with probability of 1/4. 

\item[Step 4] The change in mean CAL $Y_{it}$ and missingness indicator $M_{it}$ of each tooth are generated by regression models (\ref{regY}) and (\ref{regM}) respectively.
For the model parameters, we assume $\tau=0.85$, $\rho=0.975$, $a_0=-1$, $b_0=0.5$, $\sigma_1=0.95$ and $\sigma_0=1$ based on estimates from Bandyopadhyay and Reich \citep{Reich_Bandyopadhyay_2010}. For model (\ref{regY}), we select $\mu_i=0$ if $R_i=1$ and $\mu_i$ = 0.5, 2 or 5 if $R_i=0$ to test the proposed method. The skewness and kurtosis parameters for the error term $\epsilon_{it1}$ of model (\ref{regY}) are chosen as $\lambda=0$, $2$ and $10$, $\nu=\infty$, $10$, $8$ and $6$. This choice of parameter estimates for model (\ref{regM}) give expected proportion of available teeth around $80\%$ (i.e. $p_i\approx 0.8$) for each patient. Given the parameters of models (\ref{regY}) and (\ref{regM}), at $\lambda=10$ and $\nu=6$, the association between CAL change and missingness is around 0.44 (i.e. $c_i\approx 0.44$). 

\item[Step 5] The mean CAL change for patient $i$ is computed as $\bar{Y}_i=\dfrac{\sum_{t=1}^{28}Y_{it}(1-M_{it})}{\sum_{t=1}^{28}(1-M_{it})}$.
\end{enumerate}

Tables \ref{Table: N} - \ref{Table: N3} present a list of sample sizes calculated by the proposed method, with the corresponding Monte Carlo powers based on hypotheses 1-3, respectively. Various pairs of the skewness and kurtosis parameter $(\lambda, \nu)$ corresponding to the error $\epsilon_{it1}$ are selected. Recall, $\lambda=0$ and $\nu=\infty$ indicates normal distribution; $\lambda\neq 0$ and $\nu=\infty$ indicates skew-normal distribution; $\lambda=0$ and $\nu<\infty$ indicates $t$ distribution, and $\lambda \neq 0$ and $\nu<\infty$ indicates skew-$t$ distribution. We consider a list of effect sizes $\delta_{d_1}$ (i.e. $\mu_{d_1}$), and present their corresponding absolute values $\mid\delta_{d_1}\mid$, Monte Carlo estimates $\mid\hat{\delta}_{d_1}\mid$, standard deviations (i.e. ESD($\delta_{d_1}$)), and the Monte Carlo standard deviations (i.e., MCSD($\delta_{d_1}$)). We define $\mu_i=0$ if $R_i=1$ and $\mu_i=2$ (e.g., the first row), or 5 (e.g., the second row) if $R_i=0$ for regime 1. The absolute value of the standardized effect size is calculated as $\mid\delta_{d_1}^{\ast}\mid=\mid\delta_{d_1}/\sigma_{d_1}\mid$. We obtain small to medium (0.2-0.5) and medium to large (0.5-0.8) absolute standardized effect sizes. The results show that the Monte Carlo estimated powers ($78\%-82\%$) are close to the nominal power based on the sample size formula (\ref{smartNclustH1}), while the estimated mean and standard deviation of the effect sizes are very close to the corresponding Monte Carlo estimates.

Simulation results corresponding to hypotheses 2 and 3 are presented in Tables \ref{Table: N2} and \ref{Table: N3} respectively. They are similar to the results of Table \ref{Table: N}. The  Monte Carlo estimates of power are close to $80\%$, while the mean and standard deviation of $\hat{\delta}_{d_1-d_3}$ or $\hat{\delta}_{d_1-d_5}$ are very close to the corresponding theoretical estimates. Note, Table \ref{Table: N2} compares regimes 1 and 3. This comparison includes a pair of regimes that shares treatment `3',  where we define $\mu_i=0$ if $R_i=1$ and $\mu_i=0.5$ if $R_i=0$ for regime 1, while we define $\mu_i=0$ if $R_i=1$ and $\mu_i=2$ or 5 if $R_i=0$ for regime 3. For example, in the first two rows, $\mid\delta_{d_1-d_3}\mid=1.13$ corresponds to $\mu_i=2$ for regime 3 when $R_i=0$, while $\mid\delta_{d_1-d_3}\mid=3.38$ corresponds to $\mu_i=5$ for regime 3 when $R_i=0$. Finally, Table \ref{Table: N3} compares regimes 1 and 5, where we define $\mu_i=0$ if $R_i=1$ and $\mu_i=0.5$ if $R_i=0$ for regime 1, while we define $\mu_i=0$ if $R_i=1$ and $\mu_i=2$ or 5 if $R_i=0$ for regime 5, e.g., the first row corresponds to $\mu_i=2$ while the second row corresponds to $\mu_i=5$ for regime 5 when $R_i=0$. 

\begin{table}[H]
\caption{Estimated Sample size ($\hat{N}$) and the Monte Carlo estimated power ($\hat{P}$) at $\beta=0.2$ and $\alpha=0.05$, based on hypothesis test of $H_0:~\mu_{d_1}=0$ versus $H_1:~\mu_{d_1}\neq 0$, under different absolute value of effect size ($\mid\delta_{d_1}\mid$) with the corresponding Monte Carlo effect size ($\mid\hat{\delta}_{d_1}\mid$), standard deviation (ESD($\delta_{d_1}$)) and Monte Carlo standard deviation (MCSD($\delta_{d_1}$)), absolute value of standardized effect size ($\mid\delta_{d_1}^{\ast}\mid$) and treatments `3' response rate($\gamma^{d_1}$), skewness parameter ($\lambda$) and degree of freedom ($\nu$), given treatment `8' response rate $\gamma^{d_5}=0.5$, $\sigma_1=0.95$, $\sigma_0=1$, $\rho=0.975$, $\tau=0.85$, expected $\%$ available teeth per patient $p_i=80\%$.}
\label{Table: N}
\centering
\begin{footnotesize}
\begin{tabular}{rrrrrrrrrrr}
  \hline
 & $\mid\delta_{d_1}\mid$ & $\mid\hat{\delta}_{d_1}\mid$ & $\lambda$ & $\nu$ & $\gamma^{d_1}$ & ESD($\delta_{d_1}$) & MCSD($\delta_{d_1}$) & $\mid\delta_{d_1}^{\ast}\mid$ & $\hat{N}$ & $\hat{P}$ \\
  \hline
& 1.32 & 1.32 & 0 & Inf & 0.25 & 0.47 & 0.47 & 0.45 & 78.00 & 0.80 \\
  & 3.57 & 3.56 &   &   &   & 1.27 & 1.25 & 0.49 & 67.00 & 0.79 \\
  & 0.82 & 0.81 &   &   & 0.5 & 0.29 & 0.29 & 0.31 & 169.00 & 0.80 \\
  & 2.32 & 2.34 &   &   &   & 0.83 & 0.83 & 0.35 & 131.00 & 0.80 \\
  & 1.99 & 1.99 & 2 &   & 0.25 & 0.71 & 0.70 & 0.52 & 59.00 & 0.80 \\
  & 4.24 & 4.25 &   &   &   & 1.50 & 1.51 & 0.51 & 61.00 & 0.81 \\
  & 1.49 & 1.50 &   &   & 0.5 & 0.53 & 0.53 & 0.43 & 87.00 & 0.80 \\
  & 3.00 & 2.99 &   &   &   & 1.07 & 1.07 & 0.40 & 100.00 & 0.78 \\
  & 2.07 & 2.07 & 10 &   & 0.25 & 0.74 & 0.74 & 0.52 & 58.00 & 0.79 \\
  & 4.32 & 4.35 &   &   &   & 1.54 & 1.54 & 0.51 & 60.00 & 0.82 \\
  & 1.57 & 1.57 &   &   & 0.5 & 0.56 & 0.55 & 0.44 & 83.00 & 0.79 \\
  & 3.07 & 3.08 &   &   &   & 1.09 & 1.08 & 0.40 & 98.00 & 0.80 \\
  & 1.32 & 1.33 & 0 & 10 & 0.25 & 0.47 & 0.47 & 0.45 & 78.00 & 0.80 \\
  & 3.57 & 3.56 &   &   &   & 1.27 & 1.25 & 0.49 & 67.00 & 0.79 \\
  & 0.82 & 0.82 &   &   & 0.5 & 0.29 & 0.29 & 0.30 & 170.00 & 0.80 \\
  & 2.32 & 2.31 &   &   &   & 0.83 & 0.83 & 0.35 & 131.00 & 0.79 \\
  & 1.32 & 1.31 &   & 8 & 0.25 & 0.47 & 0.47 & 0.45 & 78.00 & 0.79 \\
  & 3.56 & 3.56 &   &   &   & 1.27 & 1.28 & 0.49 & 67.00 & 0.78 \\
  & 0.82 & 0.82 &   &   & 0.5 & 0.29 & 0.29 & 0.30 & 170.00 & 0.80 \\
  & 2.32 & 2.30 &   &   &   & 0.83 & 0.83 & 0.35 & 131.00 & 0.79 \\
  & 1.32 & 1.31 &   & 6 & 0.25 & 0.47 & 0.47 & 0.45 & 78.00 & 0.79 \\
  & 3.56 & 3.57 &   &   &   & 1.27 & 1.26 & 0.49 & 67.00 & 0.79 \\
  & 0.82 & 0.82 &   &   & 0.5 & 0.29 & 0.29 & 0.31 & 169.00 & 0.79 \\
  & 2.31 & 2.30 &   &   &   & 0.83 & 0.83 & 0.35 & 131.00 & 0.79 \\
  & 2.05 & 2.06 & 2 & 10 & 0.25 & 0.73 & 0.73 & 0.52 & 58.00 & 0.80 \\
  & 4.30 & 4.25 &   &   &   & 1.53 & 1.53 & 0.51 & 60.00 & 0.78 \\
  & 1.55 & 1.55 &   &   & 0.5 & 0.55 & 0.56 & 0.43 & 84.00 & 0.79 \\
  & 3.05 & 3.05 &   &   &   & 1.09 & 1.08 & 0.40 & 98.00 & 0.79 \\
  & 2.07 & 2.07 &   & 8 & 0.25 & 0.74 & 0.74 & 0.52 & 58.00 & 0.80 \\
  & 4.32 & 4.35 &   &   &   & 1.54 & 1.54 & 0.51 & 60.00 & 0.81 \\
  & 1.57 & 1.57 &   &   & 0.5 & 0.56 & 0.55 & 0.44 & 83.00 & 0.81 \\
  & 3.07 & 3.07 &   &   &   & 1.09 & 1.10 & 0.40 & 98.00 & 0.79 \\
  & 2.10 & 2.08 &   & 6 & 0.25 & 0.74 & 0.74 & 0.52 & 58.00 & 0.79 \\
  & 4.35 & 4.36 &   &   &   & 1.54 & 1.55 & 0.51 & 60.00 & 0.81 \\
  & 1.60 & 1.59 &   &   & 0.5 & 0.57 & 0.56 & 0.44 & 82.00 & 0.79 \\
  & 3.10 & 3.08 &   &   &   & 1.10 & 1.10 & 0.40 & 97.00 & 0.79 \\
  & 2.13 & 2.13 & 10 & 10 & 0.25 & 0.76 & 0.76 & 0.53 & 57.00 & 0.79 \\
  & 4.38 & 4.37 &   &   &   & 1.55 & 1.55 & 0.52 & 60.00 & 0.81 \\
  & 1.63 & 1.63 &   &   & 0.5 & 0.58 & 0.58 & 0.44 & 80.00 & 0.80 \\
  & 3.13 & 3.13 &   &   &   & 1.11 & 1.12 & 0.41 & 96.00 & 0.79 \\
  & 2.15 & 2.16 &   & 8 & 0.25 & 0.76 & 0.77 & 0.53 & 57.00 & 0.80 \\
  & 4.40 & 4.42 &   &   &   & 1.57 & 1.59 & 0.52 & 59.00 & 0.80 \\
  & 1.65 & 1.64 &   &   & 0.5 & 0.59 & 0.58 & 0.45 & 80.00 & 0.80 \\
  & 3.15 & 3.16 &   &   &   & 1.12 & 1.13 & 0.41 & 95.00 & 0.80 \\
  & 2.18 & 2.19 &   & 6 & 0.25 & 0.77 & 0.78 & 0.53 & 57.00 & 0.80 \\
  & 4.43 & 4.45 &   &   &   & 1.58 & 1.58 & 0.52 & 59.00 & 0.80 \\
  & 1.68 & 1.68 &   &   & 0.5 & 0.60 & 0.60 & 0.45 & 78.00 & 0.79 \\
  & 3.18 & 3.18 &   &   &   & 1.14 & 1.12 & 0.41 & 94.00 & 0.80 \\
   \hline
\end{tabular}
\end{footnotesize}
\end{table}

\begin{table}[H]
\caption{Estimated Sample size ($\hat{N}$) and the Monte Carlo estimated power ($\hat{P}$) at $\beta=0.2$ and $\alpha=0.05$, based on hypothesis test of $H_0:~\mu_{d_1}=\mu_{d_3}$ versus $H_1:~\mu_{d_1}\neq\mu_{d_3}$, under different absolute value of effect size ($\mid\delta_{d_1-d_3}\mid$) with the corresponding Monte Carlo effect size ($\mid\hat{\delta}_{d_1-d_3}\mid$), standard deviation (ESD($\delta_{d_1-d_3}$)) and Monte Carlo standard deviation (MCSD($\delta_{d_1-d_3}$)), absolute value of standardized effect size ($\mid\delta_{d_1-d_3}^{\ast}\mid$) and treatments `3' response rate($\gamma^{d_1}$), skewness parameter ($\lambda$) and degree of freedom ($\nu$), given treatment `8' response rate $\gamma^{d_5}=0.5$, $\sigma_1=0.95$, $\sigma_0=1$, $\rho=0.975$, $\tau=0.85$, expected $\%$ available teeth per patient $p_i=80\%$.}
\label{Table: N2}
\centering
\begin{footnotesize}
\begin{tabular}{rrrrrrrrrrr}
  \hline
 & $\mid\delta_{d_1-d_3}\mid$ & $\mid\hat{\delta}_{d_1-d_3}\mid$ & $\lambda$ & $\nu$ & $\gamma^{d_1}$ & ESD($\delta_{d_1-d_3}$) & MCSD($\delta_{d_1-d_3}$) & $\mid\delta_{d_1-d_3}^{\ast}\mid$ & $\hat{N}$ & $\hat{P}$ \\
  \hline
& 1.13 & 1.12 & 0 & Inf & 0.25 & 0.40 & 0.40 & 0.36 & 121.00 & 0.80 \\
& 3.38 & 3.41 &   &   &   & 1.20 & 1.20 & 0.45 & 77.00 & 0.81 \\
& 0.75 & 0.74 &   &   & 0.5 & 0.27 & 0.27 & 0.27 & 223.00 & 0.80 \\
& 2.25 & 2.26 &   &   &   & 0.80 & 0.81 & 0.33 & 141.00 & 0.80 \\
& 1.13 & 1.12 & 2 &   & 0.25 & 0.40 & 0.40 & 0.26 & 241.00 & 0.79 \\
& 3.37 & 3.40 &   &   &   & 1.20 & 1.22 & 0.39 & 105.00 & 0.80 \\
& 0.75 & 0.75 &   &   & 0.5 & 0.27 & 0.27 & 0.19 & 432.00 & 0.80 \\
& 2.25 & 2.23 &   &   &   & 0.80 & 0.80 & 0.29 & 190.00 & 0.78 \\
& 1.13 & 1.13 & 10 &   & 0.25 & 0.40 & 0.40 & 0.25 & 258.00 & 0.80 \\
& 3.37 & 3.34 &   &   &   & 1.20 & 1.18 & 0.38 & 108.00 & 0.79 \\
& 0.75 & 0.75 &   &   & 0.5 & 0.27 & 0.27 & 0.18 & 461.00 & 0.80 \\
& 2.25 & 2.22 &   &   &   & 0.80 & 0.80 & 0.28 & 195.00 & 0.79 \\
& 1.12 & 1.13 & 0 & 10 & 0.25 & 0.40 & 0.40 & 0.36 & 124.00 & 0.81 \\
& 3.37 & 3.36 &   &   &   & 1.20 & 1.19 & 0.45 & 77.00 & 0.80 \\
& 0.75 & 0.75 &   &   & 0.5 & 0.27 & 0.27 & 0.27 & 224.00 & 0.80 \\
& 2.25 & 2.25 &   &   &   & 0.80 & 0.81 & 0.33 & 141.00 & 0.80 \\
& 1.12 & 1.13 &   & 8 & 0.25 & 0.40 & 0.40 & 0.36 & 124.00 & 0.80 \\
& 3.37 & 3.39 &   &   &   & 1.20 & 1.21 & 0.45 & 77.00 & 0.79 \\
& 0.75 & 0.76 &   &   & 0.5 & 0.27 & 0.27 & 0.27 & 223.00 & 0.80 \\
& 2.25 & 2.26 &   &   &   & 0.80 & 0.79 & 0.33 & 141.00 & 0.80 \\
& 1.12 & 1.13 &   & 6 & 0.25 & 0.40 & 0.40 & 0.36 & 124.00 & 0.80 \\
& 3.37 & 3.37 &   &   &   & 1.20 & 1.21 & 0.45 & 77.00 & 0.79 \\
& 0.75 & 0.75 &   &   & 0.5 & 0.27 & 0.27 & 0.27 & 223.00 & 0.80 \\
& 2.25 & 2.25 &   &   &   & 0.80 & 0.80 & 0.33 & 142.00 & 0.80 \\
& 1.13 & 1.13 & 2 & 10 & 0.25 & 0.40 & 0.41 & 0.25 & 253.00 & 0.80 \\
& 3.38 & 3.38 &   &   &   & 1.20 & 1.22 & 0.38 & 107.00 & 0.79 \\
& 0.75 & 0.75 &   &   & 0.5 & 0.27 & 0.26 & 0.19 & 459.00 & 0.80 \\
& 2.25 & 2.24 &   &   &   & 0.80 & 0.80 & 0.28 & 195.00 & 0.79 \\
& 1.13 & 1.12 &   & 8 & 0.25 & 0.40 & 0.40 & 0.25 & 257.00 & 0.79 \\
& 3.38 & 3.36 &   &   &   & 1.20 & 1.20 & 0.38 & 108.00 & 0.79 \\
& 0.75 & 0.75 &   &   & 0.5 & 0.27 & 0.27 & 0.19 & 459.00 & 0.80 \\
& 2.25 & 2.26 &   &   &   & 0.80 & 0.80 & 0.28 & 196.00 & 0.80 \\
& 1.13 & 1.12 &   & 6 & 0.25 & 0.40 & 0.40 & 0.24 & 265.00 & 0.79 \\
& 3.37 & 3.36 &   &   &   & 1.20 & 1.19 & 0.38 & 109.00 & 0.80 \\
& 0.75 & 0.75 &   &   & 0.5 & 0.27 & 0.27 & 0.18 & 472.00 & 0.80 \\
& 2.25 & 2.26 &   &   &   & 0.80 & 0.82 & 0.28 & 199.00 & 0.80 \\
& 1.13 & 1.12 & 10 & 10 & 0.25 & 0.40 & 0.40 & 0.24 & 273.00 & 0.80 \\
& 3.37 & 3.34 &   &   &   & 1.20 & 1.20 & 0.38 & 111.00 & 0.79 \\
& 0.75 & 0.74 &   &   & 0.5 & 0.27 & 0.27 & 0.18 & 484.00 & 0.79 \\
& 2.25 & 2.25 &   &   &   & 0.80 & 0.80 & 0.28 & 202.00 & 0.80 \\
& 1.12 & 1.12 &   & 8 & 0.25 & 0.40 & 0.41 & 0.24 & 278.00 & 0.80 \\
& 3.37 & 3.38 &   &   &   & 1.20 & 1.21 & 0.38 & 112.00 & 0.80 \\
& 0.75 & 0.75 &   &   & 0.5 & 0.27 & 0.27 & 0.18 & 491.00 & 0.79 \\
& 2.25 & 2.25 &   &   &   & 0.80 & 0.81 & 0.28 & 203.00 & 0.79 \\
& 1.13 & 1.13 &   & 6 & 0.25 & 0.40 & 0.41 & 0.23 & 286.00 & 0.81 \\
& 3.38 & 3.37 &   &   &   & 1.20 & 1.22 & 0.37 & 113.00 & 0.79 \\
& 0.75 & 0.75 &   &   & 0.5 & 0.27 & 0.27 & 0.18 & 509.00 & 0.79 \\
& 2.25 & 2.23 &   &   &   & 0.80 & 0.79 & 0.28 & 206.00 & 0.79 \\
   \hline
\end{tabular}
\end{footnotesize}
\end{table}

\begin{table}[H]
\caption{Estimated Sample size ($\hat{N}$) and the Monte Carlo estimated power ($\hat{P}$) at $\beta=0.2$ and $\alpha=0.05$, based on hypothesis test of $H_0:~\mu_{d_1}=\mu_{d_5}$ versus $H_1:~\mu_{d_1}\neq\mu_{d_5}$, under different absolute value of effect size ($\mid\delta_{d_1-d_5}\mid$) with the corresponding Monte Carlo effect size ($\mid\hat{\delta}_{d_1-d_5}\mid$), standard deviation (ESD($\delta_{d_1-d_5}$)) and Monte Carlo standard deviation (MCSD($\delta_{d_1-d_5}$)), absolute value of standardized effect size ($\mid\delta_{d_1-d_5}^{\ast}\mid$) and treatments `3' response rate($\gamma^{d_1}$), skewness parameter ($\lambda$) and degree of freedom ($\nu$), given treatment `8' response rate $\gamma^{d_5}=0.5$, $\sigma_1=0.95$, $\sigma_0=1$, $\rho=0.975$, $\tau=0.85$, expected $\%$ available teeth per patient $p_i=80\%$.}
\label{Table: N3}
\centering
\begin{footnotesize}
\begin{tabular}{rrrrrrrrrrr}
  \hline
 & $\mid\delta_{d_1-d_5}\mid$ & $\mid\hat{\delta}_{d_1-d_5}\mid$ & $\lambda$ & $\nu$ & $\gamma^{d_1}$ & ESD($\delta_{d_1-d_5}$) & MCSD($\delta_{d_1-d_5}$) & $\mid\delta_{d_1-d_5}^{\ast}\mid$ & $\hat{N}$ & $\hat{P}$ \\
  \hline
& 0.63 & 0.63 & 0 & Inf & 0.25 & 0.22 & 0.22 & 0.20 & 408.00 & 0.81 \\
  & 2.12 & 2.13 &   &   &   & 0.76 & 0.76 & 0.28 & 197.00 & 0.80 \\
  & 0.75 & 0.75 &   &   & 0.5 & 0.27 & 0.27 & 0.26 & 232.00 & 0.79 \\
  & 2.25 & 2.23 &   &   &   & 0.80 & 0.80 & 0.33 & 142.00 & 0.79 \\
  & 0.62 & 0.63 & 2 &   & 0.25 & 0.22 & 0.22 & 0.14 & 789.00 & 0.80 \\
  & 2.12 & 2.12 &   &   &   & 0.76 & 0.77 & 0.24 & 264.00 & 0.80 \\
  & 0.75 & 0.75 &   &   & 0.5 & 0.27 & 0.27 & 0.19 & 448.00 & 0.80 \\
  & 2.25 & 2.25 &   &   &   & 0.80 & 0.80 & 0.29 & 192.00 & 0.80 \\
  & 0.63 & 0.63 & 10 &   & 0.25 & 0.23 & 0.23 & 0.14 & 824.00 & 0.79 \\
  & 2.12 & 2.12 &   &   &   & 0.76 & 0.74 & 0.24 & 272.00 & 0.80 \\
  & 0.75 & 0.75 &   &   & 0.5 & 0.27 & 0.27 & 0.18 & 480.00 & 0.80 \\
  & 2.25 & 2.25 &   &   &   & 0.80 & 0.80 & 0.28 & 198.00 & 0.80 \\
  & 0.62 & 0.63 & 0 & 10 & 0.25 & 0.22 & 0.22 & 0.20 & 413.00 & 0.80 \\
  & 2.13 & 2.10 &   &   &   & 0.76 & 0.75 & 0.28 & 196.00 & 0.79 \\
  & 0.75 & 0.75 &   &   & 0.5 & 0.27 & 0.27 & 0.26 & 235.00 & 0.79 \\
  & 2.25 & 2.24 &   &   &   & 0.80 & 0.80 & 0.33 & 143.00 & 0.79 \\
  & 0.63 & 0.63 &   & 8 & 0.25 & 0.22 & 0.23 & 0.20 & 411.00 & 0.80 \\
  & 2.12 & 2.14 &   &   &   & 0.76 & 0.76 & 0.28 & 197.00 & 0.80 \\
  & 0.75 & 0.75 &   &   & 0.5 & 0.27 & 0.27 & 0.26 & 235.00 & 0.79 \\
  & 2.25 & 2.24 &   &   &   & 0.80 & 0.80 & 0.33 & 143.00 & 0.79 \\
  & 0.63 & 0.63 &   & 6 & 0.25 & 0.22 & 0.22 & 0.20 & 412.00 & 0.80 \\
  & 2.12 & 2.14 &   &   &   & 0.76 & 0.76 & 0.28 & 197.00 & 0.80 \\
  & 0.75 & 0.75 &   &   & 0.5 & 0.27 & 0.27 & 0.26 & 235.00 & 0.79 \\
  & 2.25 & 2.25 &   &   &   & 0.80 & 0.80 & 0.33 & 143.00 & 0.79 \\
  & 0.62 & 0.63 & 2 & 10 & 0.25 & 0.22 & 0.22 & 0.14 & 838.00 & 0.81 \\
  & 2.13 & 2.12 &   &   &   & 0.76 & 0.77 & 0.24 & 269.00 & 0.79 \\
  & 0.75 & 0.75 &   &   & 0.5 & 0.27 & 0.27 & 0.18 & 477.00 & 0.80 \\
  & 2.25 & 2.25 &   &   &   & 0.80 & 0.80 & 0.28 & 197.00 & 0.80 \\
  & 0.63 & 0.63 &   & 8 & 0.25 & 0.22 & 0.23 & 0.14 & 835.00 & 0.80 \\
  & 2.12 & 2.13 &   &   &   & 0.76 & 0.74 & 0.24 & 273.00 & 0.81 \\
  & 0.75 & 0.75 &   &   & 0.5 & 0.27 & 0.27 & 0.18 & 478.00 & 0.80 \\
  & 2.25 & 2.25 &   &   &   & 0.80 & 0.82 & 0.28 & 198.00 & 0.79 \\
  & 0.63 & 0.63 &   & 6 & 0.25 & 0.22 & 0.22 & 0.14 & 859.00 & 0.80 \\
  & 2.13 & 2.11 &   &   &   & 0.76 & 0.76 & 0.24 & 275.00 & 0.78 \\
  & 0.75 & 0.76 &   &   & 0.5 & 0.27 & 0.27 & 0.18 & 491.00 & 0.81 \\
  & 2.25 & 2.25 &   &   &   & 0.80 & 0.80 & 0.28 & 201.00 & 0.80 \\
  & 0.63 & 0.62 & 10 & 10 & 0.25 & 0.22 & 0.22 & 0.13 & 890.00 & 0.80 \\
  & 2.12 & 2.13 &   &   &   & 0.76 & 0.76 & 0.24 & 280.00 & 0.80 \\
  & 0.75 & 0.76 &   &   & 0.5 & 0.27 & 0.27 & 0.18 & 507.00 & 0.80 \\
  & 2.25 & 2.25 &   &   &   & 0.80 & 0.81 & 0.28 & 203.00 & 0.80 \\
  & 0.63 & 0.62 &   & 8 & 0.25 & 0.22 & 0.22 & 0.13 & 900.00 & 0.79 \\
  & 2.12 & 2.14 &   &   &   & 0.76 & 0.76 & 0.24 & 282.00 & 0.80 \\
  & 0.75 & 0.75 &   &   & 0.5 & 0.27 & 0.26 & 0.17 & 519.00 & 0.80 \\
  & 2.25 & 2.25 &   &   &   & 0.80 & 0.82 & 0.28 & 205.00 & 0.79 \\
  & 0.63 & 0.63 &   & 6 & 0.25 & 0.22 & 0.22 & 0.13 & 935.00 & 0.80 \\
  & 2.13 & 2.13 &   &   &   & 0.76 & 0.76 & 0.23 & 286.00 & 0.80 \\
  & 0.75 & 0.75 &   &   & 0.5 & 0.27 & 0.27 & 0.17 & 532.00 & 0.80 \\
  & 2.25 & 2.25 &   &   &   & 0.80 & 0.80 & 0.27 & 209.00 & 0.80 \\
   \hline
\end{tabular}
\end{footnotesize}
\end{table}


\section{Implementation in R}\label{s:demo}

In this section, we demonstrate the implementation of the \texttt{R} function \texttt{SampleSize.SMARTp} for sample size calculations via a simulation study. The function is currently available for ready use via the \texttt{GitHub} link \url{https://github.com/bandyopd/SMARTp}, and forthcoming in the \texttt{CRAN} repository as a \texttt{R} package \texttt{SMARTp}. The current version of this function only considers a two-stage SMART design.

Figure \ref{fig:smart_dental} defines the SMART design.  The first three inputs of the function \texttt{SampleSize.SMARTp(mu, st1, dtr, regime, pow, b, a, rho, tau, sigma1, lambda, nu, sigma0, Num, p\_i, c\_i, a0, b0, cutoff)} are matrices, defined as:

\begin{itemize}
\item \texttt{mu}: mean matrix, where rows represent treatment paths and columns represents cluster sub-units (i.e. teeth) within a cluster (mouth),

\item \texttt{st1}: stage-1 treatment matrix, where rows represent the corresponding stage-1 treatments, the 1st column includes the numbers of treatment options for responder, the 2nd column includes the numbers of treatment options for non-responders, the 3rd column are the response rates, and the 4th column includes the row numbers of matrix `st1',

\item \texttt{dtr}: matrix of dimension (\# of DTRs X 4), the 1st column represents the DTR numbers, the 2nd column represents the corresponding treatment path numbers of responders for the corresponding DTRs in the 1st column, the third column represents the corresponding treatment path numbers of the non-responders for the corresponding DTRs in the 1st column, while the 4th column represents the corresponding initial treatment.
\end{itemize}

The \ts{regime} can be, a vector of two regime numbers if the hypothesis test is to compare regimes, or a single regime number if the hypothesis test is to detect the effect of that regime. The power, type-2 and type-1 error rates, given by \ts{pow}, \ts{b} and \ts{a} respectively, with the corresponding defaults 0.8, 0.2 and 0.05. The parameters $\tau$ and $\rho$, which quantifies the variation and association in the CAR specification of the random effect $Q_{it}$ are given by \ts{tau} and \ts{rho}, respectively, with defaults set at \ts{tau} = 0.85 and \ts{rho} = 0.975. The inputs \ts{sigma1}, \ts{lambda} and \ts{nu} define the scale ($\sigma_1$), skewness ($\lambda$) and degrees of freedom ($\nu$) parameters of the residual $\epsilon_{it1}$, which defaults to \ts{sigma1} = 0.95, \ts{lambda} = 0 and \ts{nu} = Inf. The standard deviation $\sigma_{0}$ for the residual $\epsilon_{it0}$ is given by \ts{sigma0}, whose default is \ts{sigma0} = 1. The rest of the parameters $a_0$, $b_0$ and $c_0$ from (\ref{regM}) are specified by \ts{a0}, \ts{b0} and \ts{cutoff} respectively, and their defaults are \ts{a0} = -1, \ts{b0} = 0.5 and \ts{cutoff} = 0. The user can either provide the choice of \ts{a0} and \ts{b0}, or the choice \ts{p\_i} and \ts{c\_i}, which are the expected proportion $p_i$ of available teeth for patient $i$, and the average Pearson's correlation coefficient $c_i$ between $Y_{it}$ and $M_{it0}$, averaged over the 28 teeth for patient $i$, respectively.

Monte Carlo estimates of the mean and variance of $\bar{Y}_i$ for each treatment path were obtained using \ts{Num} random samples.

The possible outputs are summarized below:
\begin{itemize}
\item \ts{N}, the calculated sample size,
\item \ts{Sigma}, the CAR covariance matrix of $Q_{it}$, i.e. $\Sigma_{28\times 28}$,
\item \ts{ybard1}, the regime mean corresponding to the 1st element of \ts{regime}, which is $\mu_{d_1}$ if, for example, \ts{regime} = c(1, 5),
\item \ts{ybard2}, the regime mean corresponding to the 2nd element of \ts{regime}, which is $\mu_{d_5}$ if, for example, \ts{regime} = c(1, 5); 0, if \ts{regime} = c(1),
\item \ts{sig.d1.sq}, N$\times$the variance of the estimated regime mean corresponding to the 1st element of \ts{regime}, which is $N\text{VAR}(\bar{Y}^{d_1})$ if, for example, \ts{regime} = c(1, 5),
\item \ts{sig.d2.sq}, N$\times$the variance of the estimated regime mean corresponding to the 2nd element of \ts{regime}, which is $N\text{VAR}(\bar{Y}^{d_5})$ if, for example, \ts{regime} = c(1, 5), or 0 if \ts{regime} = c(1),
\item \ts{sig.d1d2}, N$\times$the covariance between the estimated regime means correspond to \ts{regime}, which is $N\text{COV}(\bar{Y}^{d_1}, \bar{Y}^{d_5})$ if, for example, \ts{regime} = c(1, 5), or 0 if \ts{regime} = c(1),
\item \ts{sig.e.sq}, N$\times$the variance of the difference between the estimated regime means correspond to \ts{regime}, which is $N\text{VAR}(\bar{Y}^{d_1}-\bar{Y}^{d_5})$ if for example, \ts{regime} = c(1, 5), or $N\text{VAR}(\bar{Y}^{d_1})$ if \ts{regime} = c(1),
\item \ts{Del}, absolute value of the effect size, which is $\mid\delta_{d_1-d_5}\mid$=$\mid\mu_{d_1}$-$\mu_{d_5}\mid$ if, for example, \ts{regime} = c(1,5), or $\mid\delta_{d_1}\mid$=$\mid\mu_{d_1}\mid$, if \ts{regime} = c(1),
\item \ts{Del\_std}, absolute value of the standardized effect size, $\mid\delta_{d_1-d_5}^{\star}\mid$=$\mid\delta_{d_1-d_5}\mid/\sqrt{\text{VAR}(\bar{Y}^{d_1}-\bar{Y}^{d_5})/2}$ if, for example, \ts{regime} = c(1,5),
\item \ts{p\_st1}, randomization probability of stage-1 for each treatment path,
\item \ts{p\_st2}, randomization probability of stage-2 for each treatment path,
\item \ts{res}, a vector with binary indicators denoting responders and non-responders that corresponds to a treatment path,
\item \ts{ga}, response rates of initial treatments corresponding to each treatment path,
\item \ts{initr}, a vector with dimension as the number of treatment paths, whose elements are the corresponding row number of \ts{st1}.
\end{itemize}

\noindent In the following, we present the \ts{R} codes for the sample size calculation corresponding to the second row of Table \ref{Table: N3} in Section \ref{s:sim}.

\begin{lstlisting}[language=R]

# The SMART Design
mu=matrix(0,10,28); mu[2,]=rep(0.5,28); mu[4,]=rep(2,28);mu[7,]=rep(5,28);
st1=cbind( c(1,1), c(4,4), c(0.25, 0.5), 1:2 );
dtr=cbind( 1:8, c(rep(1,4),rep(6,4)), c(2,3,4,5,7,8,9,10), c(rep(1,4), rep(2,4)) )

## Hypothesis Test 3, with power, Type-1 and Type-2 error rates to be 80%, 5%
## and 20% respectively
regime=c(1,5); pow = 0.8; b = 1-pow; a = 0.05

## Parameter values
cutoff=0; sigma1=0.95; sigma0=1; lambda=0; nu=Inf; b0=0.5; a0=-1.0; rho=0.975;
tau=0.85; p_i=0.8027872; c_i=0.4125813

## Iteration size
Num = 1000000
\end{lstlisting}


\noindent Then, the \ts{R} codes to compute $N$, $\delta_{d_1-d_5}$, $\text{VAR}(\bar{Y}^{d_1}-\bar{Y}^{d_5})$, $\delta_{d_1-d_5}^{\star}$, $\text{VAR}(\bar{Y}^{d_1})$, $\text{VAR}(\bar{Y}^{d_5})$ and $\text{COV}(\bar{Y}^{d_1}, \bar{Y}^{d_5})$ are respectively:

\begin{lstlisting}[language=R]
SampleSize=SampleSize.SMARTp(mu=mu, st1=st1, dtr=dtr, regime=regime, pow=pow, b=b,
a=a, rho=rho, tau=tau, sigma1=sigma1, lambda=lambda, nu=nu, sigma0=sigma0,
Num=Num, p_i=p_i, c_i=c_i, cutoff=cutoff);
N=ceiling(SampleSize$N); N;
Del=SampleSize$Del; Del;
sig.e.sq=SampleSize$sig.e.sq; sig.e.sq/N;
Del_std=Del/sqrt(sig.e.sq/2); Del_std;
sig.d1.sq=SampleSize$sig.d1.sq; sig.d1.sq/N;
sig.d2.sq=SampleSize$sig.d2.sq; sig.d2.sq/N;
sig.d1d2=SampleSize$sig.d1d2; sig.d1d2/N.
\end{lstlisting}

\newpage

\section{Discussion}\label{s:disc}
This paper proposes a two-stage SMART design to study a number of DTRs for managing CP. A statistical analysis plan under this design includes hypothesis testing of detecting an effect size for either a single regime, or the difference between two regimes with or without sharing an initial treatment. This paper also develops a novel sample size calculation method, accommodating typical statistical challenges observed in CP data, such as non-Gaussianity, spatial association, and non-random misingness. To the best of our knowledge, this is the \textit{first} SMART proposal in CP research within the umbrella of \textit{precision oral health} -- a major goal in the NIH/NIDCR's Strategic Plan 2014-2019, and advances previous SMART proposals \citep{Ghosh_2016, NeCamp_2017} considered for clustered data.

An appealing feature of our method is the availability of \texttt{R} codes for implementation. However, with precision oral health as a recently emerging field, there are no real data to support the input information in the proposed sample size formula. We recommend considering plausible assumptions, such as medium effect size, conservative sample sizes, etc, to come up with the input values for implementing our proposed SMART design.  Additionally, they can be referenced by estimates from existing single-stage clinical trials. However, our experimental design, statistical analysis plan or sample size calculation can be updated or improved through data collection.

Similar to \cite{Oettingetal2011}, the proposed sample size method can also be extended to determine the optimal treatment regime. Our method can be easily updated to include more therapies (such as, various kinds of laser), and the number of treatment stages (which leads to close monitoring of CAL changes), with each stage considering more treatment types. Also, by using the Q-function approach that minimizes squared error \citep{NeCamp_2017}, or maximizing likelihood \citep{Van_Rubin_2006}, the effect size of DTRs can be adjusted by adding baseline characteristics, such as age, gender, education, oral hygiene, etc, into the regression models (\ref{regY}) or (\ref{regM}). These are important avenues for future research, and will be considered elsewhere.

\section*{Acknowledgments}
This work was supported by the grant MOE 2015-T2-2-056 from the Singapore Ministry of Education, Dr Chakraborty's start-up grant from the Duke-NUS Medical School, and the grant R01-DE024984 from the United States National Institutes of Health. The authors thank researchers at the HealthPartners Institute located at Minneapolis, Minnesota for providing the motivation, and the context behind this work. They also thank Brian Reich from the North Carolina State University for interesting discussions.

\vspace*{1pc}



\section*{Appendix} \label{s:appendix}
\addcontentsline{toc}{section}{Appendix}
\renewcommand{\thesubsection}{\Alph{subsection}}

\numberwithin{equation}{section}

\subsection{\textbf{Skew-normal distribution}} \label{SN}

\renewcommand{\theequation}{A-\arabic{equation}}
\renewcommand{\thefigure}{A-\arabic{figure}}


\setcounter{equation}{0}  
\setcounter{figure}{0}  

The statistical properties and the application of skew-normal distribution are described in \cite{Azza1996} and \cite{Azza1999} respectively. \cite{Guedes2014} presents an example of applying a regression model with skew-normal errors. Here, we present a brief introduction.

Define $Z_0\sim N(0,1)$, independent of a $m$-dimensional random variable $\boldsymbol Z=(Z_1,\ldots,Z_m)^{\top}$ with standardized normal marginals, and correlation matrix $\boldsymbol\Psi$. Suppose $\kappa_1,\ldots,\kappa_m\in(-1,1)$, define
\begin{equation}
X_j=\kappa_j\mid Z_0\mid + (1-\kappa_j^2)^{1/2}Z_j,
\end{equation}
for $j=1,\ldots,m$, where $\kappa_j=\lambda_j/(1+\lambda_j^2)^{1/2}$, such that $X_j\sim SN(\lambda_j)$, where `SN' stands for skew-normal and $\boldsymbol\lambda \in (-\infty, \infty)$ controls skewness. The probability density function of $X_j$ is $f(x_j;\lambda_j)=2\phi(x_j)\Phi(\lambda_j x_j)$,  for $-\infty<x_j<\infty$, where $\phi(\cdot)$ and $\Phi(\cdot)$ denote the density and cumulative distribution function (cdf) of $N(0,1)$, respectively. The joint density function of $X_1,\ldots,X_m$ is given by

\begin{equation}
f(\boldsymbol x; \boldsymbol\theta_x, \boldsymbol\Omega_x)=2\phi_m(\boldsymbol x; \boldsymbol\Omega_x)\Phi(\boldsymbol\theta_x^{\top} \boldsymbol x),
\end{equation}

where $\boldsymbol x =(x_1,\ldots, x_m)^{\top}$, $\phi_m(\boldsymbol x; \boldsymbol\Omega_x)$ denotes the density function of the $m$-dimension multivariate normal distribution with standardized marginals and correlation matrix $\boldsymbol\Omega_x$. We have $\boldsymbol\theta_x^{\top}=\dfrac{ \boldsymbol\lambda^{\top}\boldsymbol\Psi^{-1}\boldsymbol K^{-1} }{ ( 1 + \boldsymbol\lambda^{\top}\boldsymbol\Psi^{-1}\boldsymbol\lambda )^{1/2} }$, $K=\text{diag}((1-\kappa_{1}^2)^{1/2},\ldots,(1-\kappa_{m}^2)^{1/2})$, and $\boldsymbol\Omega_x=\boldsymbol K (\boldsymbol\Psi+\boldsymbol\lambda\boldsymbol\lambda^{\top})\boldsymbol K$.

Define $\boldsymbol Y=\boldsymbol\xi + \boldsymbol\omega \boldsymbol X$, with $\boldsymbol Y=(Y_1,\ldots Y_m)^{\top}$,
$\boldsymbol \xi=(\xi_1,\ldots \xi_m)^{\top}$ and $\boldsymbol \omega=\text{diag}(\omega_1,\ldots \omega_m)^{\top}$, where the components of $\boldsymbol \omega$ are assumed to be positive. The density function of $\boldsymbol Y$ is

\begin{equation}\label{pdfsn}
f(\boldsymbol y;\boldsymbol\xi, \boldsymbol\Omega, \boldsymbol\theta)=2\phi_m(\boldsymbol y-\boldsymbol\xi; \boldsymbol\Omega)\Phi(\boldsymbol\theta^{\top} (\boldsymbol y-\boldsymbol\xi)),
\end{equation}
where $\boldsymbol\Omega=\boldsymbol\omega\boldsymbol\Omega_x \boldsymbol\omega$ and $\boldsymbol\theta^{\top}=\boldsymbol\theta_x^{\top}\boldsymbol\omega^{-1}$. Thus, $\boldsymbol Y$ is the $m$-dimensional random variable from the SN distribution, with location $\boldsymbol\xi$, scale $\boldsymbol \omega$ and skewness $\boldsymbol\theta$, i.e. $\boldsymbol Y\sim SN_m(\boldsymbol\xi, \boldsymbol\Omega, \boldsymbol\theta)$. From (\ref{pdfsn}), we have $E(\boldsymbol Y)=\boldsymbol\xi+\boldsymbol\omega\left(\frac{2}{\pi}\right)^{1/2}\boldsymbol\kappa$, $VAR(\boldsymbol Y)=\boldsymbol\Omega-\boldsymbol\omega^2\frac{2}{\pi}\boldsymbol\kappa\boldsymbol\kappa^{\top}$, and the skewness vector $SKEW(\boldsymbol Y)=\frac{4-\pi}{2}\frac{(\boldsymbol\kappa\sqrt{2/\pi})^{3}}{(1-2\boldsymbol\kappa^2/\pi)^{3/2}}=\gamma_1$

\subsection{\textbf{Skew-$t$ distribution}} \label{ST}

\renewcommand{\theequation}{B-\arabic{equation}}
\renewcommand{\thefigure}{B-\arabic{figure}}

\setcounter{equation}{0}  
\setcounter{figure}{0}  

Skew-$t$ random variables generated from both the skew-normal and Chi-squared variables are described in \cite{Azzalini_Capitanio_2003}. Here, $\boldsymbol W\sim ST_m(\boldsymbol\xi, \boldsymbol\Omega, \boldsymbol\theta, \nu)$, such that $\boldsymbol W=\boldsymbol\xi + \boldsymbol\omega \boldsymbol X / \sqrt{V}$ and $\omega\boldsymbol X\sim SN_m(\boldsymbol 0, \boldsymbol\Omega, \boldsymbol\theta)$, where $V\sim\chi_{\nu}^{2}/\nu$ is independent of $\boldsymbol X$.


The density function of $\boldsymbol W$ is
\begin{equation}
f_W(w;\boldsymbol\xi,\boldsymbol\Omega,\boldsymbol\theta,\nu)=2t_m(w;\boldsymbol\xi,\boldsymbol\Omega,\boldsymbol\theta,\nu)T_1 \left(\boldsymbol\theta^{\top}(w-\boldsymbol\xi)\sqrt{\dfrac{\nu+m}{Q_w+\nu}}; \nu + m \right),
\end{equation}
where $Q_w=(\boldsymbol Y-\boldsymbol\xi)^{\top}\boldsymbol\Omega^{-1}(\boldsymbol Y-\boldsymbol\xi)$, $t_m(\cdot;\boldsymbol\xi,\boldsymbol\Omega,\boldsymbol\theta,\nu)$ denotes the density function of a $m$-dimensional $t$ variate with location $\boldsymbol\xi$, shape matrix $\boldsymbol\Omega$ and degrees of freedom $\nu$, while $T_1(\cdot;\nu+m)$ denotes the cdf of an univariate student's $t$ with degrees of freedom $\nu+m$. We can use the expression $\boldsymbol Y=\boldsymbol\xi + \boldsymbol\omega \boldsymbol X$ to compute the moments of Y, i.e. the $n$-th moment of $\boldsymbol Y$ is

\begin{equation}
E(\boldsymbol Y^n)=E(\boldsymbol X^n)E(V^{-n/2}),
\end{equation}
where
\begin{equation*}
E(V^{-n/2})=\dfrac{(\nu/2)^{n/2}\Gamma(\frac{1}{2}(\nu-n))}{\Gamma(\frac{1}{2}\nu)}
\end{equation*}

with $E(\boldsymbol X^n)$ given in \cite{Azza1999}. Thus, the mean and variance, are, respectively,
\begin{equation*}
E(\boldsymbol W)=\boldsymbol\xi+\boldsymbol\omega\boldsymbol\kappa(\nu/\pi)^{1/2}\dfrac{\Gamma(\frac{1}{2}(\nu-1))}{\Gamma(\frac{1}{2}\nu)},~~\nu >1,
\end{equation*}

\begin{equation*}
var(\boldsymbol W)=\dfrac{\nu}{\nu-2}\boldsymbol\Omega-\dfrac{\nu}{\pi}\left(\dfrac{\Gamma(\frac{1}{2}(\nu-1))}{\Gamma(\frac{1}{2}\nu)}\right)^2\boldsymbol\omega^2\boldsymbol\kappa \boldsymbol\kappa^{\top},~~\nu >2,
\end{equation*}

Similarly, the skewness ($\gamma_1$) and kurtosis ($\gamma_2$) for the univariate cases are
\begin{equation*}
\gamma_1=\mu\left[ \dfrac{\nu(3-\kappa^2)}{\nu-3}- \dfrac{3\nu}{\nu-2} +2\mu^2 \right]\left[ \dfrac{\nu}{\nu-2}-\mu^2 \right]^{-3/2},~~\nu >3,
\end{equation*}
\begin{equation*}
\gamma_2=\left[\dfrac{3\nu^2}{(\nu-2)(\nu-4)} - \dfrac{4\mu^2\nu(3-\kappa^2)}{\nu-3} + \dfrac{6\mu^2\nu}{\nu-2} -3\mu^4 \right]\left[ \dfrac{\nu}{\nu-2}-\mu^2 \right]^{-2}-3,~~\nu >4,
\end{equation*}
where $\mu=\kappa\sqrt{\dfrac{\nu}{\pi}}\dfrac{\Gamma(\frac{1}{2}(\nu-1))}{\Gamma(\frac{1}{2}\nu)}$.

\subsection{\textbf{Sample size formula derivation}}\label{SamFormDerApp}

\renewcommand{\theequation}{C-\arabic{equation}}
\renewcommand{\thefigure}{C-\arabic{figure}}

\setcounter{equation}{0}  
\setcounter{figure}{0}  

The covariance between $Y_{it}$ and $M_{it0}$ is
\begin{align*}
&\text{COV}(\mu_i+Q_{it}+\epsilon_{it1}, a_0+b_0Q_{it}+\epsilon_{it0})\\
=&\text{COV}(Q_{it}, b_0 Q_{it})\\
=&b_0\text{VAR}(Q_{it})\\
=&b_0\Sigma_{tt},
\end{align*}
where $\Sigma_{tt}$ is the $t^{\text{th}}$ diagonal elements of the covariance matrix $\boldsymbol\Sigma_{28\times 28}$. The Pearson correlation coefficient is
\begin{align*}
c_{it}&=\frac{ b_0\text{VAR}(Q_{it}) }{ \sqrt{ \text{VAR}(\mu_i+Q_{it}+\epsilon_{it1}) \text{VAR}(a_0+b_0Q_{it}+\epsilon_{it0}) } }\\
&=\frac{ b_0\text{VAR}(Q_{it}) }{ \sqrt{ (\text{VAR}(Q_{it})+\text{VAR}(\epsilon_{it1}))(b_0^2 \text{VAR}(Q_{it})+\text{VAR}(\epsilon_{it0})) } },\\
\end{align*}
where $\text{VAR}(Q_{it})$ = $\Sigma_{tt}$, $\text{VAR}(\epsilon_{it0})$ = $\sigma_0^2$ and $\text{VAR}(\epsilon_{it1})$ = $\frac{\sigma_1^2\nu}{\nu - 2}$ - $\frac{\nu}{\pi} \left[ \frac{ \Gamma( 0.5(\nu - 1) ) }{ \Gamma( 0.5\nu ) } \right]^2 \frac{ \sigma_1^2\lambda^2}{1+\lambda^2}$ if $\nu<\infty$, otherwise $\text{VAR}(\epsilon_{it1})$=$\sigma_1^2$ - $\frac{2}{\pi} \frac{ \sigma_1^2\lambda^2}{1+\lambda^2}$. Now, we derive an expression for $p_i$, i.e.,

\begin{align*}
p_i=&1-E\left( \frac{\sum_{t=1}^{28}M_{it}}{28} \right)\\
=&1-\frac{1}{28}\sum_{t=1}^{28}E\left[ E( M_{it} \mid Q_{it}, \epsilon_{it0} ) \right]\\
=&1-\frac{1}{28}\sum_{t=1}^{28}E\left[ E( I(a_0 + b_0 Q_{it} + \epsilon_{it0}>c_0) \mid Q_{it}, \epsilon_{it0} ) \right] \\
=&1-\frac{1}{28}\sum_{t=1}^{28}E\left[ I(a_0 + b_0 Q_{it} + \epsilon_{it0}>c_0) \right] \\
=&1-\frac{1}{28}\sum_{t=1}^{28}\text{Pr}(a_0 + b_0 Q_{it} + \epsilon_{it0}>c_0)\\
=&1-\frac{1}{28}\sum_{t=1}^{28}\text{Pr}\left(z>\frac{ c_0-E(a_0 + b_0 Q_{it} + \epsilon_{it0}) }{ \sqrt{ \text{VAR}(a_0 + b_0 Q_{it} + \epsilon_{it0}) } }\right)\\
=&1-\frac{1}{28}\sum_{t=1}^{28}\left[ 1-\Phi\left( \frac{ c_0-a_0 }{ \sqrt{b_0^2\Sigma_{tt} + \sigma_0^2} } \right) \right]
\end{align*}
where $\Phi(\cdot)$ is the cdf of $z\sim$N(0,1). Now,

\begin{align*}
E(\bar{Y}^{d_1})=&E\left( \dfrac{ I( A_{i1}=a_{1i}^{d_1}, A_{i2}=[a_{2i}^{d_1 R}]^{R_i}[a_{2i}^{d_1 NR}]^{1-R_i} ) }{ \pi_{1i}^{d_1} [\pi_{2i}^{d_1 R}]^{R_i}[\pi_{2i}^{d_1 NR}]^{1-R_i} }\bar{Y}_{i} \right)\\
=&E\left[ E\left( \dfrac{ I( A_{i1}=a_{1i}^{d_1}, A_{i2}=[a_{2i}^{d_1 R}]^{R_i}[a_{2i}^{d_1 NR}]^{1-R_i} ) }{ \pi_{1i}^{d_1} [\pi_{2i}^{d_1 R}]^{R_i}[\pi_{2i}^{d_1 NR}]^{1-R_i} }\bar{Y}_{i} \mid R_i \right)\right]\\
=&\gamma^{d_1}\mu_{d_1 R}+(1-\gamma^{d_1})\mu_{d_1 NR}.
\end{align*}
In a similar way, we have $E(\bar{Y}^{d_3})=\gamma^{d_3}\mu_{d_3 R}+(1-\gamma^{d_3})\mu_{d_3 NR}$.

According to variance decomposition, the right side of (\ref{varybard1}) is the sum of two components, which are
\begin{equation*}
E[ V( \dfrac{ I( A_{i1}=a_{1i}^{d_1}, A_{i2}=[a_{2i}^{d_1 R}]^{R_i}[a_{2i}^{d_1 NR}]^{1-R_i} ) }{ \pi_{1i}^{d_1} [\pi_{2i}^{d_1 R}]^{R_i}[\pi_{2i}^{d_1 NR}]^{1-R_i} }\bar{Y}_{i} \mid R_i ) ]
\end{equation*}
and
\begin{equation*}
V[ E( \dfrac{ I( A_{i1}=a_{1i}^{d_1}, A_{i2}=[a_{2i}^{d_1 R}]^{R_i}[a_{2i}^{d_1 NR}]^{1-R_i} ) }{ \pi_{1i}^{d_1} [\pi_{2i}^{d_1 R}]^{R_i}[\pi_{2i}^{d_1 NR}]^{1-R_i} }\bar{Y}_{i} \mid R_i ) ].
\end{equation*}
The first component is
\begin{align*}
&E[ V( \dfrac{ I( A_{i1}=a_{1i}^{d_1}, A_{i2}=[a_{2i}^{d_1 R}]^{R_i}[a_{2i}^{d_1 NR}]^{1-R_i} ) }{ \pi_{1i}^{d_1} [\pi_{2i}^{d_1 R}]^{R_i}[\pi_{2i}^{d_1 NR}]^{1-R_i} }\bar{Y}_{i} \mid R_i ) ]\\
=&\sum_{R_i=0}^{1}V(\dfrac{ I( A_{i1}=a_{1i}^{d_1}, A_{i2}=[a_{2i}^{d_1 R}]^{R_i}[a_{2i}^{d_1 NR}]^{1-R_i} ) }{ \pi_{1i}^{d_1} [\pi_{2i}^{d_1 R}]^{R_i}[\pi_{2i}^{d_1 NR}]^{1-R_i} }\bar{Y}_{i} \mid R_i)Pr(R_i),
\end{align*}
while the second component is
\begin{align*}
&V[ E( \dfrac{ I( A_{i1}=a_{1i}^{d_1}, A_{i2}=[a_{2i}^{d_1 R}]^{R_i}[a_{2i}^{d_1 NR}]^{1-R_i} ) }{ \pi_{1i}^{d_1} [\pi_{2i}^{d_1 R}]^{R_i}[\pi_{2i}^{d_1 NR}]^{1-R_i} }\bar{Y}_{i} \mid R_i ) ]\\
=&\sum_{R_i=0}^{1}E^{2}(\dfrac{ I( A_{i1}=a_{1i}^{d_1}, A_{i2}=[a_{2i}^{d_1 R}]^{R_i}[a_{2i}^{d_1 NR}]^{1-R_i} ) }{ \pi_{1i}^{d_1} [\pi_{2i}^{d_1 R}]^{R_i}[\pi_{2i}^{d_1 NR}]^{1-R_i} }\bar{Y}_{i} \mid R_i)Pr(R_i)\\
&-( \sum_{R_i=0}^{1}E(\dfrac{ I( A_{i1}=a_{1i}^{d_1}, A_{i2}=[a_{2i}^{d_1 R}]^{R_i}[a_{2i}^{d_1 NR}]^{1-R_i} ) }{ \pi_{1i}^{d_1} [\pi_{2i}^{d_1 R}]^{R_i}[\pi_{2i}^{d_1 NR}]^{1-R_i} }\bar{Y}_{i} \mid R_i)Pr(R_i) )^{2},
\end{align*}
based on the formulae  $E(X)=E[E(X\mid Y)]$ and $V(X)=E(X^2)-E^{2}(X)$. We have $Pr(R_i=1)=\gamma^{d_1}$ or $Pr(R_i=0)=1-\gamma^{d_1}$, and
\begin{align*}
&V( \dfrac{ I( A_{i1}=a_{1i}^{d_1}, A_{i2}=[a_{2i}^{d_1 R}] ) }{ \pi_{1i}^{d_1} [\pi_{2i}^{d_1 R}] }\bar{Y}_{i} \mid R_i=1 )\\
=&E( \dfrac{ I( A_{i1}=a_{1i}^{d_1}, A_{i2}=[a_{2i}^{d_1 R}] ) }{ (\pi_{1i}^{d_1} [\pi_{2i}^{d_1 R}])^2 }\bar{Y}_{i}^{2} ) -
E^{2}( \dfrac{ I( A_{i1}=a_{1i}^{d_1}, A_{i2}=[a_{2i}^{d_1 R}] ) }{ \pi_{1i}^{d_1} [\pi_{2i}^{d_1 R}] }\bar{Y}_{i} )\\
=&\dfrac{1}{\pi_{1i}^{d_1} [\pi_{2i}^{d_1 R}]}E_{d_1 R}(\bar{Y}_{i}^{2})-E_{d_1 R}^{2}(\bar{Y}_{i})\\
=&\dfrac{1}{\pi_{1i}^{d_1} [\pi_{2i}^{d_1 R}]}( \sigma_{d_{1}R}^{2} + ( 1-\pi_{1i}^{d_1} [\pi_{2i}^{d_1 R}] )\mu_{d_{1}R}^{2} ),
\end{align*}
with $\mu_{d_{1}R}$, the expectation of $\bar{Y}_i$ from $d_1$, with $R_i=1$, given by
\begin{align*}
&E_{d_1 R}(\bar{Y}_{i})\\
=&E_{d_1 R}\left(E_{d_1 R}\left(\dfrac{\sum_{t=1}^{28}Y_{it}(1-M_{it})}{\sum_{t=1}^{28}(1-M_{it})}\mid \boldsymbol Q_{i}, \boldsymbol\epsilon_{i0}\right)\right)\\
=&E_{d_1 R}\left(E_{d_1 R}\left(\dfrac{\sum_{t=1}^{28}(\mu_i + Q_{it}+\epsilon_{it1})(1-I(a_{0}+b_{0}Q_{it}+\epsilon_{it0}>0))}{\sum_{t=1}^{28}(1-I(a_{0}+b_{0}Q_{it}+\epsilon_{it0}>0))}\mid \boldsymbol Q_{i}, \boldsymbol\epsilon_{i0}\right)\right)\\
=&E\left(\dfrac{\sum_{t=1}^{28}(\mu_{i}\mid_{A_{i3}=1, R_i=1} + Q_{it}+E(\epsilon_{it1}))(1-I(a_{0}+b_{0}Q_{it}+\epsilon_{it0}>0))}{\sum_{t=1}^{28}(1-I(a_{0}+b_{0}Q_{it}+\epsilon_{it0}>0))}\right)\\
=&\int_{\boldsymbol Q_{i}}\int_{\boldsymbol \epsilon_{i}}\dfrac{\sum_{t=1}^{28}(\mu_{i}\mid_{A_{i3}=1, R_i=1} + Q_{it}+E(\epsilon_{it1}))(1-I(a_{0}+b_{0}Q_{it}+\epsilon_{it0}>0))}{\sum_{t=1}^{28}(1-I(a_{0}+b_{0}Q_{it}+\epsilon_{it0}>0))}f(\boldsymbol Q_{i})f(\boldsymbol\epsilon_{i})d\boldsymbol\epsilon_{i}d\boldsymbol Q_{i},
\end{align*}
where $f(\boldsymbol Q_{i})$ and $f(\boldsymbol\epsilon_{i})$ are the density functions for $\boldsymbol Q_{i}$ and $\boldsymbol \epsilon_{i}$ respectively. Also, $\sigma_{d_{1}R}^{2}$, the variance of $\bar{Y}_i$ that is from $d_1$, with $R_i=1$, can be written as
\begin{align*}
V_{d_1 R}(\bar{Y}_{i})=E_{d_1 R}(V_{d_1 R}(\bar{Y}_{i}\mid \boldsymbol Q_{i}, \boldsymbol\epsilon_{i0})) + V_{d_1 R}(E_{d_1 R}(\bar{Y}_{i}\mid \boldsymbol Q_{i}, \boldsymbol\epsilon_{i0})),
\end{align*}
where
\begin{align*}
&E_{d_1 R}(V_{d_1 R}(\bar{Y}_{i}\mid \boldsymbol Q_{i}, \boldsymbol\epsilon_{i0}))\\
=&E_{d_1 R}\left(V_{d_1 R}\left(\dfrac{\sum_{t=1}^{28}Y_{it}(1-M_{it})}{\sum_{t=1}^{28}(1-M_{it})}\mid \boldsymbol Q_{i}, \boldsymbol\epsilon_{i0}\right)\right) \\
=&E\left(V\left( \dfrac{\sum_{t=1}^{28}(\mu_i\mid_{A_{i3}=1, R_i=1} + Q_{it}+\epsilon_{it1})(1-I(a_{0}+b_{0}Q_{it}+\epsilon_{it0}>0))}{\sum_{t=1}^{28}(1-I(a_{0}+b_{0}Q_{it}+\epsilon_{it0}>0))} \right)\right)\\
=&E\left( \dfrac{\sum_{t=1}^{28}[1-I(a_{0}+b_{0}Q_{it}+\epsilon_{it0}>0)]\sigma_1^2}{\left(\sum_{t=1}^{28}[1-I(a_{0}+b_{0}Q_{it}+\epsilon_{it0}>0)]\right)^2} \right)\\
=&\int_{\boldsymbol Q_{i}}\int_{\boldsymbol \epsilon_{i}}\dfrac{\sum_{t=1}^{28}[1-I(a_{0}+b_{0}Q_{it}+\epsilon_{it0}>0)]\sigma_1^2}{\left(\sum_{t=1}^{28}[1-I(a_{0}+b_{0}Q_{it}+\epsilon_{it0}>0)]\right)^2}f(\boldsymbol Q_{i})f(\boldsymbol\epsilon_{i})d\boldsymbol\epsilon_{i}d\boldsymbol Q_{i}
\end{align*}
and
\begin{align*}
&V_{d_1 R}(E_{d_1 R}(\bar{Y}_{i}\mid \boldsymbol Q_{i}, \boldsymbol\epsilon_{i0}))\\
=&V_{d_1 R}\left(E_{d_1 R}\left(\dfrac{\sum_{t=1}^{28}Y_{it}(1-M_{it})}{\sum_{t=1}^{28}(1-M_{it})}\mid \boldsymbol Q_{i}, \boldsymbol\epsilon_{i0}\right)\right)\\
=&V\left(E\left( \dfrac{\sum_{t=1}^{28}(\mu_i\mid_{A_{i3}=1, R_i=1} + Q_{it}+\epsilon_{it1})(1-I(a_{0}+b_{0}Q_{it}+\epsilon_{it0}>0))}{\sum_{t=1}^{28}(1-I(a_{0}+b_{0}Q_{it}+\epsilon_{it0}>0))} \right)\right)\\
=&V\left( \dfrac{\sum_{t=1}^{28}(\mu_i\mid_{A_{i3}=1, R_i=1} + Q_{it}+E(\epsilon_{it1}))(1-I(a_{0}+b_{0}Q_{it}+\epsilon_{it0}>0))}{\sum_{t=1}^{28}(1-I(a_{0}+b_{0}Q_{it}+\epsilon_{it0}>0))} \right)\\
=&E\left[\left( \dfrac{\sum_{t=1}^{28}(\mu_i\mid_{A_{i3}=1, R_i=1} + Q_{it}+E(\epsilon_{it1}))(1-I(a_{0}+b_{0}Q_{it}+\epsilon_{it0}>0))}{\sum_{t=1}^{28}(1-I(a_{0}+b_{0}Q_{it}+\epsilon_{it0}>0))} \right)^2\right]\\
&-\left[ E\left( \dfrac{\sum_{t=1}^{28}(\mu_i\mid_{A_{i3}=1, R_i=1} + Q_{it}+E(\epsilon_{it1}))(1-I(a_{0}+b_{0}Q_{it}+\epsilon_{it0}>0))}{\sum_{t=1}^{28}(1-I(a_{0}+b_{0}Q_{it}+\epsilon_{it0}>0))} \right)\right]^2\\
=&\int_{\boldsymbol Q_{i}}\int_{\boldsymbol \epsilon_{i}}\left( \dfrac{\sum_{t=1}^{28}(\mu_i\mid_{A_{i3}=1, R_i=1} + Q_{it}+E(\epsilon_{it1}))(1-I(a_{0}+b_{0}Q_{it}+\epsilon_{it0}>0))}{\sum_{t=1}^{28}(1-I(a_{0}+b_{0}Q_{it}+\epsilon_{it0}>0))} \right)^2 f(\boldsymbol Q_{i})f(\boldsymbol\epsilon_{i})d\boldsymbol\epsilon_{i}d\boldsymbol Q_{i}\\
-&\left[ \int_{\boldsymbol Q_{i}}\int_{\boldsymbol \epsilon_{i}} \dfrac{\sum_{t=1}^{28}(\mu_i\mid_{A_{i3}=1, R_i=1} + Q_{it}+E(\epsilon_{it1}))(1-I(a_{0}+b_{0}Q_{it}+\epsilon_{it0}>0))}{\sum_{t=1}^{28}(1-I(a_{0}+b_{0}Q_{it}+\epsilon_{it0}>0))} f(\boldsymbol Q_{i})f(\boldsymbol\epsilon_{i})d\boldsymbol\epsilon_{i}d\boldsymbol Q_{i} \right]^2.
\end{align*}

Similarly, we have

\begin{align*}
V( \dfrac{ I( A_{i1}=a_{1i}^{d_1}, A_{i2}=[a_{2i}^{d_1 NR}] ) }{ \pi_{1i}^{d_1} [\pi_{2i}^{d_1 NR}] }\bar{Y}_{i} \mid R_i=0 ) =
\dfrac{1}{\pi_{1i}^{d_1} [\pi_{2i}^{d_1 NR}]}( \sigma_{d_{1}NR}^{2} + ( 1-\pi_{1i}^{d_1} [\pi_{2i}^{d_1 NR}] )\mu_{d_{1}NR}^{2} )
\end{align*}, where $\mu_{d_{1}NR}^{2}$ and $\sigma_{d_{1}NR}^{2}$ are the expectation and variance of $\bar{Y}_i$ from $d_1$ with $R_i=0$.

Therefore, the second component is
\begin{align*}
&V[ E( \dfrac{ I( A_{i1}=a_{1i}^{d_1}, A_{i2}=[a_{2i}^{d_1 R}]^{R_i}[a_{2i}^{d_1 NR}]^{1-R_i} ) }{ \pi_{1i}^{d_1} [\pi_{2i}^{d_1 R}]^{R_i}[\pi_{2i}^{d_1 NR}]^{1-R_i} }\bar{Y}_{i} \mid R_i ) ]\\
=&\gamma^{d_1}\mu_{d_{1}R}^{2} + (1-\gamma^{d_1})\mu_{d_{1}NR}^{2} -(\gamma^{d_1}\mu_{d_{1}R} + (1-\gamma^{d_1})\mu_{d_{1}NR})^2\\
=&\gamma^{d_1}(1-\gamma^{d_1})(\mu_{d_{1}R}- \mu_{d_{1}NR})^{2}.
\end{align*}

Thus, the variance formula (\ref{varybard1}) is
\begin{align*}
&V(\bar{Y}^{d_1}) \\
=\dfrac{1}{N} & \{ \dfrac{\gamma^{d_1}}{\pi_{1i}^{d_1} [\pi_{2i}^{d_1 R}]}( \sigma_{d_{1} R}^{2} + ( 1-\pi_{1i}^{d_1} [\pi_{2i}^{d_1 R}] )\mu_{d_{1} R}^{2} ) +
\dfrac{1-\gamma^{d_1}}{\pi_{1i}^{d_1} [\pi_{2i}^{d_1 NR}]}( \sigma_{d_{1}NR}^{2} + ( 1-\pi_{1i}^{d_1} [\pi_{2i}^{d_1 NR}] )\mu_{d_{1}NR}^{2} )+ \\
&\gamma^{d_1}(1-\gamma^{d_1})(\mu_{d_{1}R}- \mu_{d_{1}NR})^{2} \}
\end{align*}


while $V(\bar{Y}^{d_3})$ is
\begin{align*}
&\dfrac{1}{N}\{ \dfrac{\gamma^{d_3}}{\pi_{1i}^{d_3} [\pi_{2i}^{d_3 R}]}( \sigma_{d_{3} R}^{2} + ( 1-\pi_{1i}^{d_3} [\pi_{2i}^{d_3 R}] )\mu_{d_{3} R}^{2} ) +
\dfrac{1-\gamma^{d_3}}{\pi_{1i}^{d_3} [\pi_{2i}^{d_3 NR}]}( \sigma_{d_{3}NR}^{2} + ( 1-\pi_{1i}^{d_3} [\pi_{2i}^{d_3 NR}] )\mu_{d_{3}NR}^{2} )+ \\
&\gamma^{d_3}(1-\gamma^{d_3})(\mu_{d_{3}R}- \mu_{d_{3}NR})^{2} \}.
\end{align*}

The variance of the difference between $d_1$ and $d_3$ is
\begin{equation}\label{varexact}
V(\bar{Y}^{d_1}-\bar{Y}^{d_3})=V(\bar{Y}^{d_1})+V(\bar{Y}^{d_3})-2\text{COV}(\bar{Y}^{d_1}, \bar{Y}^{d_3}).
\end{equation}
The covariance between $\bar{Y}^{d_1}$ and $\bar{Y}^{d_3}$ is
\begin{align*}
&\text{COV}\left(\bar{Y}^{d_1}, \bar{Y}^{d_3}\right)\\
=&\dfrac{1}{N^2}\text{COV}\left(\sum_{i=1}^{N}W_{i}^{d_1}\bar{Y}_i, \sum_{i=1}^{N}W_{i}^{d_3}\bar{Y}_i\right)\\
=&\dfrac{1}{N^2}\text{COV}\left(\sum_{i=1}^{N}W_{i}^{d_1}\bar{Y}_i (R_i+(1-R_i)), \sum_{i=1}^{N}W_{i}^{d_3}\bar{Y}_i (R_i+(1-R_i))\right)\\
=&\dfrac{1}{N^2}\text{COV}\left(\sum_{i=1}^{N}W_{i}^{d_1}\bar{Y}_i R_i+W_{i}^{d_1}\bar{Y}_i(1-R_i), \sum_{i=1}^{N}W_{i}^{d_3}\bar{Y}_i R_i+W_{i}^{d_3}\bar{Y}_i(1-R_i)\right)\\
=&\dfrac{1}{N^2}\text{COV}\left(\sum_{i=1}^{N}W_{i}^{d_1}\bar{Y}_i R_i+ \sum_{i=1}^{N}W_{i}^{d_1}\bar{Y}_i (1-R_i), \sum_{i=1}^{N}W_{i}^{d_3}\bar{Y}_i R_i + \sum_{i=1}^{N}W_{i}^{d_3}\bar{Y}_i (1-R_i)\right)\\
=&\dfrac{1}{N^2}\text{COV}\left(\sum_{i=1}^{N}W_{i}^{d_1}\bar{Y}_i R_i, \sum_{i=1}^{N}W_{i}^{d_3}\bar{Y}_i R_i \right)+\dfrac{1}{N^2}\text{COV}\left(\sum_{i=1}^{N}W_{i}^{d_1}\bar{Y}_i R_i, \sum_{i=1}^{N}W_{i}^{d_3}\bar{Y}_i (1-R_i) \right)\\
&+\dfrac{1}{N^2}\text{COV}\left(\sum_{i=1}^{N}W_{i}^{d_1}\bar{Y}_i (1-R_i), \sum_{i=1}^{N}W_{i}^{d_3}\bar{Y}_i R_i \right)+\dfrac{1}{N^2}\text{COV}\left(\sum_{i=1}^{N}W_{i}^{d_1}\bar{Y}_i (1-R_i), \sum_{i=1}^{N}W_{i}^{d_3}\bar{Y}_i (1-R_i) \right)\\
=&\dfrac{1}{N^2}\left[\sum_{i=1}^{N}\text{COV}\left(W_{i}^{d_1}\bar{Y}_i R_i, W_{i}^{d_3}\bar{Y}_i R_i\right) + \sum_{i\neq j}\text{COV}\left(W_{i}^{d_1}\bar{Y}_i R_i, W_{j}^{d_3}\bar{Y}_j R_j\right)\right]\\
&+\dfrac{1}{N^2}\left[\sum_{i=1}^{N}\text{COV}\left(W_{i}^{d_1}\bar{Y}_i R_i, W_{i}^{d_3}\bar{Y}_i (1-R_i)\right) + \sum_{i\neq j}\text{COV}\left(W_{i}^{d_1}\bar{Y}_i R_i, W_{j}^{d_3}\bar{Y}_j (1-R_j)\right)\right]\\
&+\dfrac{1}{N^2}\left[\sum_{i=1}^{N}\text{COV}\left(W_{i}^{d_1}\bar{Y}_i (1-R_i), W_{i}^{d_3}\bar{Y}_i R_i\right) + \sum_{i\neq j}\text{COV}\left(W_{i}^{d_1}\bar{Y}_i (1-R_i), W_{j}^{d_3}\bar{Y}_j R_j\right)\right]\\
&+\dfrac{1}{N^2}\left[\sum_{i=1}^{N}\text{COV}\left(W_{i}^{d_1}\bar{Y}_i (1-R_i), W_{i}^{d_3}\bar{Y}_i (1-R_i)\right) + \sum_{i\neq j}\text{COV}\left(W_{i}^{d_1}\bar{Y}_i (1-R_i), W_{j}^{d_3}\bar{Y}_j (1-R_j)\right)\right]\\
=&\dfrac{1}{N}\text{COV}\left(W_{i}^{d_1}\bar{Y}_i R_i, W_{i}^{d_3}\bar{Y}_i R_i \right)+\dfrac{1}{N}\text{COV}\left(W_{i}^{d_1}\bar{Y}_i R_i, W_{i}^{d_3}\bar{Y}_i (1-R_i) \right)\\
&+\dfrac{1}{N}\text{COV}\left(W_{i}^{d_1}\bar{Y}_i (1-R_i), W_{i}^{d_3}\bar{Y}_i R_i \right)+\dfrac{1}{N}\text{COV}\left(W_{i}^{d_1}\bar{Y}_i (1-R_i), W_{i}^{d_3}\bar{Y}_i (1-R_i) \right),\\
\end{align*}
where
\begin{align*}
&\text{COV}\left(W_{i}^{d_1}\bar{Y}_i R_i, W_{i}^{d_3}\bar{Y}_i R_i \right)\\
=&E(W_{i}^{d_1}\bar{Y}_i R_i W_{i}^{d_3}\bar{Y}_i R_i)-E(W_{i}^{d_1}\bar{Y}_i R_i)E(W_{i}^{d_3}\bar{Y}_i R_i)\\
=&E[E(W_{i}^{d_1}\bar{Y}_i R_i W_{i}^{d_3}\bar{Y}_i R_i)\mid R_i]-E[E(W_{i}^{d_1}\bar{Y}_i R_i)\mid R_i ]E[E(W_{i}^{d_3}\bar{Y}_i R_i)\mid R_i]\\
=&\dfrac{\gamma^{d_1}}{\pi_{1i}^{d_1} \pi_{2i}^{d_1 R}}E_{d_1 R}(\bar{Y}_i^2) \\
&-\gamma^{d_1}\gamma^{d_3}E_{d_1 R}(\bar{Y}_i)E_{d_3 R}(\bar{Y}_i)\\
=&\dfrac{\gamma^{d_1}}{\pi_{1i}^{d_1} \pi_{2i}^{d_1 R}}\left[ \sigma_{d_1 R}^{2} + \mu_{d_1 R}^{2} \right] -\gamma^{d_1}\gamma^{d_3} \mu_{d_1 R}\mu_{d_3 R},\\
\end{align*}

Similarly,
\begin{align*}
&\text{COV}\left(W_{i}^{d_1}\bar{Y}_i R_i, W_{i}^{d_3}\bar{Y}_i (1-R_i) \right)\\
=&E[E(W_{i}^{d_1}\bar{Y}_i R_i W_{i}^{d_3}\bar{Y}_i (1-R_i))\mid R_i]-E[E(W_{i}^{d_1}\bar{Y}_i R_i)\mid R_i ]E[E(W_{i}^{d_3}\bar{Y}_i (1-R_i))\mid R_i]\\
=&-\gamma^{d_1}(1-\gamma^{d_3})\mu_{d_1 R}\mu_{d_3 NR},\\
\end{align*}
\begin{align*}
&\text{COV}\left(W_{i}^{d_1}\bar{Y}_i (1-R_i), W_{i}^{d_3}\bar{Y}_i R_i \right)\\
=&E[E(W_{i}^{d_1}\bar{Y}_i (1-R_i) W_{i}^{d_3}\bar{Y}_i R_i)\mid R_i]-E[E(W_{i}^{d_1}\bar{Y}_i (1-R_i))\mid R_i ]E[E(W_{i}^{d_3}\bar{Y}_i R_i)\mid R_i]\\
=&-\gamma^{d_3}(1-\gamma^{d_1})\mu_{d_1 NR}\mu_{d_3 R},\\
\end{align*}
\begin{align*}
&\text{COV}\left(W_{i}^{d_1}\bar{Y}_i (1-R_i), W_{i}^{d_3}\bar{Y}_i (1-R_i) \right)\\
=&E[E(W_{i}^{d_1}\bar{Y}_i (1-R_i) W_{i}^{d_3}\bar{Y}_i (1-R_i))\mid R_i]-E[E(W_{i}^{d_1}\bar{Y}_i (1-R_i))\mid R_i ]E[E(W_{i}^{d_3}\bar{Y}_i (1-R_i))\mid R_i]\\
=&-(1-\gamma^{d_1})(1-\gamma^{d_3})\mu_{d_1 NR}\mu_{d_3 NR}.\\
\end{align*}

Thus, $\text{COV}\left(\bar{Y}^{d_1}, \bar{Y}^{d_3}\right)$ is
\begin{align*}
&\text{COV}\left(\bar{Y}^{d_1}, \bar{Y}^{d_3}\right)\\
=&\dfrac{1}{N}\{\dfrac{\gamma^{d_1}}{\pi_{1i}^{d_1} \pi_{2i}^{d_1 R}}(\sigma_{d_1 R}^{2} + \mu_{d_1 R}^{2} ) -\gamma^{d_1}\gamma^{d_3} \mu_{d_1 R}\mu_{d_3 R}\\
&-\gamma^{d_1}(1-\gamma^{d_3})\mu_{d_1 R}\mu_{d_3 NR}\\
&-\gamma^{d_3}(1-\gamma^{d_1})\mu_{d_1 NR}\mu_{d_3 R}\\
&-(1-\gamma^{d_1})(1-\gamma^{d_3})\mu_{d_1 NR}\mu_{d_3 NR}\}.\\
\end{align*}

Therefore, the variance of regime means differences between $d_1$ and $d_3$ is
\begin{align*}
&V(\bar{Y}^{d_1}-\bar{Y}^{d_3})\\
=\dfrac{1}{N}&\{\dfrac{\gamma^{d_1}}{\pi_{1i}^{d_1} [\pi_{2i}^{d_1 R}]}( \sigma_{d_{1} R}^{2} + ( 1-\pi_{1i}^{d_1} [\pi_{2i}^{d_1 R}] )\mu_{d_{1} R}^{2} ) +  \\
&\dfrac{1-\gamma^{d_1}}{\pi_{1i}^{d_1} [\pi_{2i}^{d_1 NR}]}( \sigma_{d_{1}NR}^{2} + ( 1-\pi_{1i}^{d_1} [\pi_{2i}^{d_1 NR}] )\mu_{d_{1}NR}^{2} )+ \\
&\gamma^{d_1}(1-\gamma^{d_1})(\mu_{d_{1}R}- \mu_{d_{1}NR})^{2}+\\
&\dfrac{\gamma^{d_3}}{\pi_{1i}^{d_3} [\pi_{2i}^{d_3 R}]}( \sigma_{d_{3} R}^{2} + ( 1-\pi_{1i}^{d_3} [\pi_{2i}^{d_3 R}] )\mu_{d_{3} R}^{2} ) +  \\
&\dfrac{1-\gamma^{d_3}}{\pi_{1i}^{d_3} [\pi_{2i}^{d_3 NR}]}( \sigma_{d_{3}NR}^{2} + ( 1-\pi_{1i}^{d_3} [\pi_{2i}^{d_3 NR}] )\mu_{d_{3}NR}^{2} )+ \\
&\gamma^{d_3}(1-\gamma^{d_3})(\mu_{d_{3}R}- \mu_{d_{3}NR})^{2}\\
&-2[\dfrac{\gamma^{d_1}}{\pi_{1i}^{d_1}\pi_{2i}^{d_1 R}}(\sigma_{d_1 R}^{2} + \mu_{d_1 R}^{2} ) -\gamma^{d_1}\gamma^{d_3} \mu_{d_1 R}\mu_{d_3 R}\\
&-\gamma^{d_1}(1-\gamma^{d_3})\mu_{d_1 R}\mu_{d_3 NR}\\
&-\gamma^{d_3}(1-\gamma^{d_1})\mu_{d_1 NR}\mu_{d_3 R}\\
&-(1-\gamma^{d_1})(1-\gamma^{d_3})\mu_{d_1 NR}\mu_{d_3 NR} ] \}.\\
\end{align*}

Similarly, we can also derive the expression for $\text{COV}\left(\bar{Y}^{d_1}, \bar{Y}^{d_5}\right)$, where $d_1$ and $d_5$ does not share an initial treatment, i.e.
\begin{align*}
&\text{COV}\left(\bar{Y}^{d_1}, \bar{Y}^{d_5}\right)\\
=&\dfrac{1}{N}\{-\gamma^{d_1}\gamma^{d_5} \mu_{d_1 R}\mu_{d_5 R}\\
&-\gamma^{d_1}(1-\gamma^{d_5})\mu_{d_1 R}\mu_{d_5 NR}\\
&-\gamma^{d_5}(1-\gamma^{d_1})\mu_{d_1 NR}\mu_{d_5 R}\\
&-(1-\gamma^{d_1})(1-\gamma^{d_5})\mu_{d_1 NR}\mu_{d_5 NR}\}.\\
\end{align*}

\subsection{\textbf{Proof of Theorem \ref{theo1}}}\label{ThePro}

\renewcommand{\theequation}{D-\arabic{equation}}
\renewcommand{\thefigure}{D-\arabic{figure}}

\setcounter{equation}{0}  
\setcounter{figure}{0}  

\noindent \textbf{Proof:} \\
The proof ofconsistency requires the result of strong law of large numbers, such that \\ $\hat{\delta}_{d_1-d_3}$ = $\bar{Y}_{d_1}-\bar{Y}_{d_3}$=$\dfrac{1}{N}\sum_{i=1}^{N}W^{d_1}_{i}\bar{Y}_{i}$-$\dfrac{1}{N}\sum_{i=1}^{N}W^{d_3}_{i}\bar{Y}_{i}\rightarrow\mu_{d_10}-\mu_{d_30}$, almost surely, and uniformly for $\delta_{d_1-d_3}\in\Theta$ as $N\rightarrow\infty$ and $\delta_{(d_1-d_3)0}$ being the unique expected value of $\hat{\delta}_{d_1-d_3}$ due to Assumption 2. To prove the asymptotic normality result, we have
\begin{align*}
&\sqrt{N}(\hat{\delta}_{d_1-d_3}-\delta_{(d_1-d_3)0})\\
=&\sqrt{N}\left[\dfrac{1}{N}\sum_{i=1}^{N}W^{d_1}_{i}\bar{Y}_{i}-\dfrac{1}{N}\sum_{i=1}^{N}W^{d_3}_{i}\bar{Y}_{i}-(\mu_{d_10}-\mu_{d_30})\right]
\end{align*}
with $E\left(\dfrac{1}{N}\sum_{i=1}^{N}W^{d_1}_{i}\bar{Y}_{i}\right)=\mu_{d_10}$, $E\left(\dfrac{1}{N}\sum_{i=1}^{N}W^{d_3}_{i}\bar{Y}_{i}\right)=\mu_{d_30}$, $E\left(\hat{\delta}_{d_1-d_3}\right)=\delta_{(d_1-d_3)0}$ and $\text{var}(\hat{\delta}_{d_1-d_3})=2\sigma_{d_1-d_3}^2/N$. Thus, by the central limit theorem, $\sqrt{N}(\hat{\delta}_{d_1-d_3}-\delta_{(d_1-d_3)0})$ converges in distribution to $N(0,2\sigma^2_{d_1-d_3})$.




\begin{singlespace}
\bibliographystyle{rss}
\bibliography{spatialSMART}

\begin{thebibliography}{46}
\expandafter\ifx\csname natexlab\endcsname\relax\def\natexlab#1{#1}\fi
\expandafter\ifx\csname url\endcsname\relax
  \def\url#1{\texttt{#1}}\fi
\expandafter\ifx\csname urlprefix\endcsname\relax\def\urlprefix{URL}\fi

\bibitem[{Aparecida~Guedes et~al.(2014)Aparecida~Guedes, Rossi, Tozzo~Martins,
  Janeiro and Pedroza~Carneiro}]{Guedes2014}
Aparecida~Guedes, T., Rossi, R.~M., Tozzo~Martins, A.~B., Janeiro, V. and
  Pedroza~Carneiro, J.~W. (2014) {Applying regression models with skew-normal
  errors to the height of bedding plants of Stevia rebaudiana (Bert) Bertoni}.
\newblock \textit{{Acta Scientiarum. Technology}}, \textbf{36}, 463--468.

\bibitem[{Azarpazhooh et~al.(2010)Azarpazhooh, Shah, Tenenbaum and
  Goldberg}]{azarpazhooh2010effect}
Azarpazhooh, A., Shah, P.~S., Tenenbaum, H.~C. and Goldberg, M.~B. (2010) The
  effect of photodynamic therapy for periodontitis: a systematic review and
  meta-analysis.
\newblock \textit{Journal of Periodontology}, \textbf{81}, 4--14.

\bibitem[{Azzalini and Capitanio(1999)}]{Azza1999}
Azzalini, A. and Capitanio, A. (1999) Statistical applications of the
  multivariate skew normal distribution.
\newblock \textit{Journal of the Royal Statistical Society: Series B
  (Statistical Methodology)}, \textbf{61}, 579--602.

\bibitem[{Azzalini and
  Capitanio(2003{\natexlab{a}})}]{azzalini2003distributions}
--- (2003{\natexlab{a}}) {Distributions generated by perturbation of symmetry
  with emphasis on a multivariate skew-$t$ distribution}.
\newblock \textit{Journal of the Royal Statistical Society: Series B
  (Statistical Methodology)}, \textbf{65}, 367--389.

\bibitem[{Azzalini and Capitanio(2003{\natexlab{b}})}]{Azzalini_Capitanio_2003}
--- (2003{\natexlab{b}}) Distributions generated by perturbation of symmetry
  with emphasis on a multivariate skew-t distribution.
\newblock \textit{Journal of the Royal Statistical Society: Series B
  (Statistical Methodology)}, \textbf{65}, 367--–389.

\bibitem[{Azzalini and Dalla~Valle(1996)}]{Azza1996}
Azzalini, A. and Dalla~Valle, A. (1996) The multivariate skew-normal
  distribution.
\newblock \textit{Biometrika}, \textbf{83}, 715--726.

\bibitem[{Beck et~al.(2001)Beck, Elter, Heiss, Couper, Mauriello and
  Offenbacher}]{beck2001relationship}
Beck, J.~D., Elter, J.~R., Heiss, G., Couper, D., Mauriello, S.~M. and
  Offenbacher, S. (2001) {Relationship of periodontal disease to carotid artery
  intima-media wall thickness: the atherosclerosis risk in communities (ARIC)
  study}.
\newblock \textit{Arteriosclerosis, Thrombosis, and Vascular Biology},
  \textbf{21}, 1816--1822.

\bibitem[{Breslin et~al.(1998)Breslin, Sobell, Sobell, Cunningham, Sdao-Jarvie
  and Borsoi}]{Breslinetal1999}
Breslin, F.~C., Sobell, M.~B., Sobell, L.~C., Cunningham, J.~A., Sdao-Jarvie,
  K. and Borsoi, D. (1998) Problem drinkers: evaluation of a stepped-care
  approach.
\newblock \textit{Journal of Substance Abuse}, \textbf{10}, 217--232.

\bibitem[{Brooner and Kidorf(2002)}]{Brooner2002}
Brooner, R. and Kidorf, M. (2002) Using behavioral reinforcement to improve
  methadone treatment participation.
\newblock \textit{Science and Practice Perspectives}, \textbf{1}, 38--47.

\bibitem[{Cao et~al.(2009)Cao, Tsiatis and Davidian}]{Cao_etal_2009}
Cao, W., Tsiatis, A. and Davidian, M. (2009) Improving efficiency and
  robustness of the doubly robust estimator for a population mean with
  incomplete data.
\newblock \textit{Biometrika}, \textbf{96}, 723--734.

\bibitem[{Chakraborty and Moodie(2013)}]{Chakraborty_2013}
Chakraborty, B. and Moodie, E. (2013) \textit{Statistical Methods for Dynamic
  Treatment Regimes}.
\newblock New York: Springer.

\bibitem[{Chakraborty et~al.(2010)Chakraborty, Murphy and
  Strecher}]{Chakraborty_etal_2010}
Chakraborty, B., Murphy, S. and Strecher, V. (2010) Inference for non-regular
  parameters in optimal dynamic treatment regimes.
\newblock \textit{Statistical Methods in Medical Research}, \textbf{19},
  317–--343.

\bibitem[{Eke et~al.(2012)Eke, Page, Wei, Thornton-Evans and
  Genco}]{Eke_etal_2012}
Eke, P., Page, R., Wei, L., Thornton-Evans, G. and Genco, R. (2012) {Update of
  the Case Definitions for Population-Based Surveillance of Periodontitis}.
\newblock \textit{Journal of Periodontology}, \textbf{83}, 1449--1454.

\bibitem[{Fernandes et~al.(2009)Fernandes, Wiegand, Salinas, Grossi, Sanders,
  Lopes-Virella and Slate}]{fernandes2009periodontal}
Fernandes, J.~K., Wiegand, R.~E., Salinas, C.~F., Grossi, S.~G., Sanders,
  J.~J., Lopes-Virella, M.~F. and Slate, E.~H. (2009) {Periodontal disease
  status in Gullah African Americans with Type-2 diabetes living in South
  Carolina}.
\newblock \textit{Journal of Periodontology}, \textbf{80}, 1062--1068.

\bibitem[{Garcia et~al.(2013)Garcia, Kuska and Somerman}]{garcia2013expanding}
Garcia, I., Kuska, R. and Somerman, M. (2013) {Expanding the Foundation for
  Personalized Medicine: Implications and Challenges for Dentistry}.
\newblock \textit{Journal of Dental Research Clinical Research Supplement},
  \textbf{92}, 3S--10S.

\bibitem[{Ghosh et~al.(2016)Ghosh, Cheung and Chakraborty}]{Ghosh_2016}
Ghosh, P., Cheung, Y. and Chakraborty, B. (2016) {Sample size calculations for
  clustered SMART designs}.
\newblock In \textit{{Adaptive Treatment Strategies in Practice: Planning
  Trials and Analyzing Data for Personalized Medicine}} (eds. M.~Kosorok and
  E.~Moodie), chap.~5, 55--68. Philadelphia, PA: ASA-SIAM Statistics and
  Applied Probability Series.

\bibitem[{Glasgow et~al.(1989)Glasgow, Engel and D’Lugoff}]{Glasgowetal1989}
Glasgow, M., Engel, B. and D’Lugoff, B. (1989) A controlled study of a
  standardized behavioural stepped treatment for hypertension.
\newblock \textit{Psychosomatic Medicine}, \textbf{51}, 10--26.

\bibitem[{Grossi et~al.(1997)Grossi, Skrepcinski, DeCaro, Robertson, Ho,
  Dunford and Genco}]{grossi1997treatment}
Grossi, S.~G., Skrepcinski, F.~B., DeCaro, T., Robertson, D.~C., Ho, A.~W.,
  Dunford, R.~G. and Genco, R.~J. (1997) Treatment of periodontal disease in
  diabetics reduces glycated hemoglobin.
\newblock \textit{Journal of Periodontology}, \textbf{68}, 713--719.

\bibitem[{Herrera(2016)}]{herrera2016scaling}
Herrera, D. (2016) Scaling and root planning is recommended in the nonsurgical
  treatment of chronic periodontitis.
\newblock \textit{Journal of Evidence-Based Dental Practice}, \textbf{16},
  56--58.

\bibitem[{John et~al.(2017)John, Michalowicz, Kotsakis and
  Chu}]{John_etal_2017}
John, M., Michalowicz, B., Kotsakis, G. and Chu, H. (2017) {Network
  meta-analysis of studies included in the Clinical Practice Guideline on the
  nonsurgical treatment of chronic periodontitis}.
\newblock \textit{Journal of Clinic Peridontology}, \textbf{44}, 603--611.

\bibitem[{Lavori and Dawson(2004)}]{Lavorietal2000}
Lavori, P.~W. and Dawson, R. (2004) {Dynamic treatment regimes: Practical
  design considerations}.
\newblock \textit{Clinical Trials}, \textbf{1}, 9--20.

\bibitem[{Lei et~al.(2012)Lei, Nahum-Shani, Lynch, Oslin and
  Murphy}]{lei2012smart}
Lei, H., Nahum-Shani, I., Lynch, K., Oslin, D. and Murphy, S.~A. (2012) {A
  ``SMART'' Design for Building Individualized Treatment Sequences}.
\newblock \textit{Annual Review of Clinical Psychology}, \textbf{8}, 21--48.

\bibitem[{Liu et~al.(1999)Liu, Hou, Wong and Lan}]{liu1999comparison}
Liu, C.-M., Hou, L.-T., Wong, M.-Y. and Lan, W.-H. (1999) {Comparison of Nd:
  YAG laser versus scaling and root planing in periodontal therapy}.
\newblock \textit{Journal of Periodontology}, \textbf{70}, 1276--1282.

\bibitem[{Murphy(2003)}]{Murphy2003}
Murphy, S.~A. (2003) Optimal dynamic treatment regimes.
\newblock \textit{Journal of the Royal Statistical Society: Series B
  (Statistical Methodology)}, \textbf{65}, 331--366.

\bibitem[{Murphy(2005)}]{Murphy_2005}
--- (2005) An experimental design for the development of adaptive treatment
  strategies.
\newblock \textit{Statistics in Medicine}, \textbf{24}, 1455--1481.

\bibitem[{Murphy et~al.(2001)Murphy, van~der Laan, Robins and
  Group}]{Murphyetal2001}
Murphy, S.~A., van~der Laan, M.~J., Robins, J.~M. and Group, C. P. P.~R. (2001)
  Marginal mean models for dynamic regimes.
\newblock \textit{Journal of the American Statistical Association},
  \textbf{96}, 1410--1423.

\bibitem[{Murphy and McKay(2004)}]{MurphyMcKay2003}
Murphy, S.~A. and McKay, J.~R. (2004) {Adaptive treatment strategies: An
  emerging approach for improving treatment effectiveness}.
\newblock \textit{Clinical Science}, \textbf{12}, 7--13.

\bibitem[{NeCamp et~al.(2017)NeCamp, Kilbourne and Almirall1}]{NeCamp_2017}
NeCamp, T., Kilbourne, A. and Almirall1, D. (2017) {Comparing cluster-level
  dynamic treatment regimens using sequential, multiple assignment, randomized
  trials: Regression estimation and sample size considerations}.
\newblock \textit{Statistical Methods in Medical Research}, \textbf{26},
  1572--1589.

\bibitem[{Nicholls(2003)}]{Nicholls_2003}
Nicholls, C. (2003) {Periodontal disease incidence, progression and rate of
  tooth loss in a general dental practice: The results of a 12-year
  retrospective analysis of patient's clinical records}.
\newblock \textit{British Dental Journal}, \textbf{194}, 485--488.

\bibitem[{Oetting et~al.(2011)Oetting, Levy, Weiss and
  Murphy}]{Oettingetal2011}
Oetting, A., Levy, J., Weiss, R. and Murphy, S. (2011) Statistical methodology
  for a smart design in the development of adaptive treatment strategies.
\newblock In \textit{P. E. Shrout, K. M. Keyes, K. Ornstein (Eds.), Causality
  and psychopathology: Finding the determinants of disorders and their cures},
  179--205. Arlington, VA: American Psychiatric Publishing.

\bibitem[{Porteous and Rowe(2014)}]{porteous2014adjunctive}
Porteous, M.~S. and Rowe, D.~J. (2014) Adjunctive use of the diode laser in
  non-surgical periodontal therapy: Exploring the controversy.
\newblock \textit{{The Journal of Dental Hygiene}}, \textbf{88}, 78--86.

\bibitem[{Reich et~al.(2013)Reich, Bandyopadhyay and Bondell}]{Reich_etal_2013}
Reich, B., Bandyopadhyay, D. and Bondell, H. (2013) {A Nonparametric Spatial
  Model for Periodontal Data with Nonrandom Missingness}.
\newblock \textit{Journal of the American Statistical Association},
  \textbf{108}, 820--831.

\bibitem[{Reich and Bandyopadhyay(2010)}]{Reich_Bandyopadhyay_2010}
Reich, B.~J. and Bandyopadhyay, D. (2010) A latent factor model for spatial
  data with informative missingness.
\newblock \textit{The Annals of Applied Statistics}, \textbf{4}, 439--459.

\bibitem[{Robins et~al.(1994{\natexlab{a}})Robins, Rotnitzky and
  Zhao}]{Robins_etal_1994}
Robins, J., Rotnitzky, A. and Zhao, L. (1994{\natexlab{a}}) Estimation of
  regression coefficients when some regressors are not always observed.
\newblock \textit{Journal of the American Statistical Association},
  \textbf{89}, 846--866.

\bibitem[{Robins(2004)}]{Robins2004}
Robins, J.~M. (2004) {Optimal Structural Nested Models for Optimal Sequential
  Decisions}.
\newblock In \textit{{Proceedings of the Second Seattle Symposium in
  Biostatistics: Analysis of Correlated Data}} (eds. D.~Lin and P.~Heagerty),
  vol. 179 of \textit{{Lecture Notes in Statistics}}, chap.~11, 189--326. New
  York: Springer.

\bibitem[{Robins et~al.(1994{\natexlab{b}})Robins, Rotnitzky and
  Zhao}]{robins1994estimation}
Robins, J.~M., Rotnitzky, A. and Zhao, L.~P. (1994{\natexlab{b}}) Estimation of
  regression coefficients when some regressors are not always observed.
\newblock \textit{Journal of the American Statistical Association},
  \textbf{89}, 846--866.

\bibitem[{Sgolastra et~al.(2012)Sgolastra, Petrucci, Gatto and
  Monaco}]{sgolastra2012efficacy}
Sgolastra, F., Petrucci, A., Gatto, R. and Monaco, A. (2012) {Efficacy of
  Er:YAG laser in the treatment of chronic periodontitis: systematic review and
  meta-analysis}.
\newblock \textit{Lasers in Medical Science}, \textbf{27}, 661--673.

\bibitem[{Smiley et~al.(2015)Smiley, Tracy, Abt, Michalowicz, John, Gunsolley,
  Cobb, Rossmann, Harrel, Forrest, Hujoel, Noraian, Greenwell, Frantsve-Hawley,
  Estrich and Hanson}]{Smiley_etal_2015}
Smiley, C., Tracy, S., Abt, E., Michalowicz, B., John, M., Gunsolley, J., Cobb,
  C., Rossmann, J., Harrel, S., Forrest, J., Hujoel, P., Noraian, K.,
  Greenwell, H., Frantsve-Hawley, J., Estrich, C. and Hanson, N. (2015)
  {Systematic review and meta-analysis on the nonsurgical treatment of chronic
  periodontitis by means of scaling and root planing with or without adjuncts}.
\newblock \textit{Journal of American Dental Association}, \textbf{146},
  508--524.

\bibitem[{Thornton-Evans et~al.(2013)Thornton-Evans, Eke, Wei, Palmer, Moeti,
  Hutchins and Borrell}]{Thornton_etal_2013}
Thornton-Evans, G., Eke, P., Wei, L., Palmer, A., Moeti, R., Hutchins, S. and
  Borrell, L. (2013) {Periodontitis among adults aged $\geq$30 years - United
  States, 2009-2010}.
\newblock \textit{{Centers for Disease Control and Prevention. Morbidity and
  mortality weekly report}}, \textbf{62}, 129--135.

\bibitem[{Untzer et~al.(2001)Untzer, Katon, Williams, Callahan, Harpole,
  Hunkeler, Hoffing, Arean, Hegel, Schoenbaum, Oishi and
  Langston}]{Untzeretal2001}
Untzer, J., Katon, W., Williams, J., Callahan, C., Harpole, L., Hunkeler, E.,
  Hoffing, M., Arean, P., Hegel, M., Schoenbaum, M., Oishi, S. and Langston, C.
  (2001) Improving primary care for depression in late life: the design of a
  multicenter randomized trial.
\newblock \textit{Medical Care}, \textbf{39}, 785–--799.

\bibitem[{Van Der~Laan and Rubin(2006)}]{Van_Rubin_2006}
Van Der~Laan, M. and Rubin, D. (2006) {Targeted Maximum Likelihood Learning}.
\newblock \textit{Working Paper Series Working Paper 213}, U.C. Berkeley
  Division of Biostatistics.

\bibitem[{Vonesh et~al.(2006)Vonesh, Greene and Schluchter}]{vonesh2006shared}
Vonesh, E.~F., Greene, T. and Schluchter, M.~D. (2006) Shared parameter models
  for the joint analysis of longitudinal data and event times.
\newblock \textit{Statistics in Medicine}, \textbf{25}, 143--163.

\bibitem[{Wang et~al.(2009)Wang, Zhou, Zhang, Zhang, Song, Hu and
  Wang}]{wang2009periodontal}
Wang, Z., Zhou, X., Zhang, J., Zhang, L., Song, Y., Hu, F.~B. and Wang, C.
  (2009) Periodontal health, oral health behaviours, and chronic obstructive
  pulmonary disease.
\newblock \textit{Journal of Clinical Periodontology}, \textbf{36}, 750--755.

\bibitem[{Wiebe and Putnins(2000)}]{wiebe2000periodontal}
Wiebe, C.~B. and Putnins, E.~E. (2000) {The Periodontal Disease Classification
  System of the American Academy of Periodontology -- An Update}.
\newblock \textit{Journal of the Canadian Dental Association}, \textbf{66},
  594--597.

\bibitem[{Workgroup(2011)}]{perio2011AAP}
Workgroup, A. (2011) {American Academy of Periodontology Statement on the
  efficacy of lasers in the non-surgical treatment of inflammatory periodontal
  disease}.
\newblock \textit{Journal of Periodontology}, \textbf{82}, 513--514.

\bibitem[{Zhao et~al.(2014)Zhao, Yin, Tao, Nie, Tang and Zhu}]{zhao2014er}
Zhao, Y., Yin, Y., Tao, L., Nie, P., Tang, Y. and Zhu, M. (2014) {Er:YAG laser
  versus scaling and root planing as alternative or adjuvant for chronic
  periodontitis treatment: a systematic review}.
\newblock \textit{Journal of Clinical Periodontology}, \textbf{41}, 1069--1079.

\end{thebibliography}
\end{singlespace}


\end{document}